\documentclass[manuscript]{acmart}
\AtBeginDocument{%
  \providecommand\BibTeX{{%
    \normalfont B\kern-0.5em{\scshape i\kern-0.25em b}\kern-0.8em\TeX}}}

\setcopyright{acmcopyright}
\copyrightyear{2024}
\acmYear{2024}
\acmDOI{XXXXXXX.XXXXXXX}

\acmConference[CSCW '25]{ACM Conference on Computer-Supported Cooperative Work and Social Computing}{October 18 — 22, 2025}{Bergen, Norway}
\acmPrice{15.00}
\acmISBN{978-1-4503-XXXX-X/18/06}

\usepackage{enumitem}
\usepackage{xcolor,colortbl}
\usepackage{multirow}
\usepackage{multicol}
\usepackage{array}
\usepackage{siunitx}
\usepackage{graphicx}
\usepackage{ragged2e}
\usepackage{soul}
\usepackage{lscape}
\usepackage{graphics}
\usepackage{supertabular}
\usepackage{amsmath}

\usepackage{dblfloatfix}

\usepackage{xspace}

\renewcommand{\paragraph}[1]{\vspace{0.2em}\noindent\textbf{\textit{#1}}\hspace*{.3em}}

\definecolor{navy}{RGB}{0,0,180}
\definecolor{maroon}{RGB}{128,0,0}
\definecolor{forest}{RGB}{0,100,0}
\newcommand{\revision}[1]{{\color{black}#1\normalfont}}
\usepackage[normalem]{ulem}

\begin{document}

\title[Authenticity in Co-Writing with LLM]{"It was 80\% me, 20\% AI":\\ Seeking Authenticity in Co-Writing with Large Language Models}

% "Because I did most of the work": Seeking Authenticity in Creative Writing with Large-Language Models

\author{Angel Hsing-Chi Hwang}
\email{angel.hwang@usc.edu}
\affiliation{%
  \institution{University of Southern California}
  \city{Los Angeles}
  \state{CA}
  \country{USA}
}

\author{Q. Vera Liao}
\affiliation{%
  \institution{Microsoft Research Montréal}
  \country{Canada}}
\email{veraliao@microsoft.com}

\author{Su Lin Blodgett}
\affiliation{%
  \institution{Microsoft Research Montréal}
  \country{Canada}
}
\email{sulin.blodgett@microsoft.com}

\author{Alexandra Olteanu}
\affiliation{%
 \institution{Microsoft Research Montréal}
 \country{Canada}}
\email{alexandra.olteanu@microsoft.com}

\author{Adam Trischler}
\affiliation{%
  \institution{Microsoft Research Montréal}
  \country{Canada}}

 \renewcommand{\shortauthors}{Hwang et al.}

\begin{abstract}
Given the rising proliferation and diversity of AI writing assistance tools, especially those powered by large language models (LLMs), both writers and readers may have concerns about the impact of these tools on the authenticity of writing work.
We examine whether and how writers want to preserve their authentic voice when co-writing with AI tools and whether personalization of AI writing support could help achieve this goal.
We conducted semi-structured interviews with 19 professional writers, during which they co-wrote with both personalized and non-personalized AI writing-support tools.
We supplemented writers’ perspectives with opinions from 30 avid readers about the written work co-produced with AI collected through an online survey.
Our findings illuminate conceptions of authenticity in human-AI co-creation, which focus more on the process and experience of constructing creators’ authentic selves. While writers reacted positively to personalized AI writing tools, they believed the form of personalization needs to target writers’ growth and go beyond the phase of text production. Overall, readers’ responses showed less concern about human-AI co-writing. Readers could not distinguish AI-assisted work, personalized or not, from writers’ solo-written work and showed positive attitudes toward writers experimenting with new technology for creative writing. 
\end{abstract}



% \keywords{datasets, neural networks, gaze detection, text tagging}

%   \includegraphics[width=0.9\linewidth]{fig/data_stream_cite.jpg}
%   \caption{\todo{TO DO} \cite{mohr2017personal}.}
% %   \Description{Enjoying the baseball game from the third-base seats. Ichiro Suzuki preparing to bat.}
%   \label{fig:data_stream_cite}


\settopmatter{printfolios=true} % add page number

\maketitle

\section{Introduction}

From text suggestion \cite{Mina_Lee_CoAuthor} and summarization \cite{AIwriting_auto_summary} to style transformation \cite{AIwriting_Tool_arbitrary_style_transfer}, metaphor generation \cite{AIwriting_metaphorian}, and information synthesis \cite{news_summ_eval_2022}, burgeoning applications of artificial intelligence (AI) for text production seem to be rapidly reshaping writing experiences and practices, especially with the recent high-profile releases of large language models (LLMs). 
Consequently, there are also concerns that vast transformations of the writer economy are likely underway~\cite{Merali_economic_productivity_2024, MJE_2024, GPTs_are_GPTs_2024, Brookings_FoW}.
Within such a climate, seeking and preserving \textit{\textbf{authenticity}}---as a cornerstone for all forms of creation---in writing content co-created with AI is likely to become an increasingly complicated yet critical matter for writers.

Indeed, existing literature has pointed to the importance of understanding authenticity for several reasons:
From writers' perspectives, authenticity often determines the value of their work, which co-writing with AI might potentially threaten~\cite{HAIC_authenticity_Social_Dynamics_Gero}.
Moreover, writing serves as the medium for writers to connect with their audiences, and authentic expression contributes to the soundness of such bonds~\cite{writer-reader_Candlin_Hyland_1999, writer-reader_sharing_words, writer-reader_writing_for_readers}.
A deeper understanding of authenticity also facilitates discussions around ownership of work \cite{HAIC_authenticity_AI_Ghostwriter_Effect} and relevant practices such as declaring authorship, regulating copyright, detecting plagiarism, and commissioning writers' work. 
% -- as human-AI co-creation becomes commonplace \tocite{}.
Recent work on AI use for writing has begun to explore relevant constructs, such as ownership, authorship, and agency~\cite{mieczkowski2022ai, HAIC_authenticity_AI_Ghostwriter_Effect, HAIC_authenticity_choice_over_control, HAIC_tool_agency}, 
%such as agency AI writing assistance \tocite{} and perceived authorship of AI-generated text \tocite{}.
but \revision{a more comprehensive understanding of views surrounding authenticity in human-AI co-creation remains elusive.}
% Meanwhile, public discussions reveal growing concerns about the potential threats of AI assistance tools to creators' value and authenticity in their work \cite{news_Lee_2023}.
Though public discussions reveal growing concerns from writers about the impact of AI on their work and profession~\cite{news_Lee_2023}, it remains unclear whether and how they would like to preserve elements of authenticity in writing.

Meanwhile, personalized AI applications---including \textit{personalized AI writing assistance}---are becoming more common and readily accessible~\cite{personalization_CIKM_workshop, personalization_NLG, personalization_social_media_misinfo, personalization_user_centric_TTT}. This is believed to be especially promising with LLMs, \revision{which can be prompted or fine-tuned to generate a more specific form or style of text,
%through natural language based prompting techniques, 
allowing people with various degrees of AI/ML expertise to experiment with personalizing or steering text generation.} 
%With emergent features (e.g., fine-tuning, prompting, in-context learning) powered by large-language models (LLM), even users without much AI/ML expertise could easily custom-build their AI writing assistance. 
For instance, a user could try to personalize AI writing suggestions to simulate their own writing style by specifying the characteristics of their desired style in the prompt or by providing a few of their own writing samples (i.e., in-context learning~\cite{brown2020language}).
But can such personalization be sufficient to help preserve writers' authentic voices in writing?
While some recent research suggests personalized AI might add little to writers' perceived ownership of their co-created writing work with AI~\cite{HAIC_authenticity_AI_Ghostwriter_Effect}, the potentials of and concerns about personalizing AI writing suggestions to support authenticity remain largely under-explored.

% Vera's comments:
% Another research gap/motivation is to study personalization. I think we can highlight LLMs’ opportunities for easy personalization, possible benefits of personalization for authenticity, and post it as an open question, e.g. “ different from previous generation of AI or automated writing support tools, LLMs allow for easy adaptation (e.g. through fine-tuning, prompting including incontext learning). This opens up possibilities for writer to personalize the model and better inject their own styles in the generation. However, it remains an open question whether writers and readers perceive writing done with personalized laguage models to be more authentic [while a recent study found that personalization had little effect on related constructs such as perceived author ownership (https://arxiv.org/pdf/2303.03283.pdf [AI Ghostwriter Effect])---this paper is very relevant, but my understanding is they tested actually personalized v.s. placebo personalized, not exactly personalized v.s. non-personalized

%Given these various motivations, the present work takes 
In this work, we take a closer look at authenticity in writing from both writers' and readers' perspectives. 
We focus on what writers seek for authenticity as new practices of co-writing with AI emerge and whether personalization could support their goals.
\revision{Furthermore, as personalized AI tools become readily available, we seek to understand the possible impact of personalization on writers' ability to express and preserve their authentic voices in writing.}
Specifically, we ask:
\begin{itemize}
    \item RQ1: How do writers and readers conceptualize authenticity in the context of human-AI co-writing?
    \item RQ2: Based on their conceptions of authenticity, do writers want to preserve their authentic voices in writing, and if so, how?
    \item RQ3: Can personalized AI writing assistance support authenticity and help preserve writers' authentic voices (if desired) in writing, and if so, how?
\end{itemize}
%\textbf{\textit{How do writers and readers conceptualize authenticity in the context of human-AI co-creation?}}
We examined these questions first through semi-structured interviews with 19 professional writers across various literature genres.
During the interviews, writer participants reflected on their conceptions of authenticity in writing \textit{and} co-writing with AI through situated experiences. 
Specifically, they engaged in writing with both generic and personalized AI writing assistance powered by a state-of-the-art large language model (GPT-4).
We then complemented writers' perspectives with those from avid readers through an online survey (N = 30), which allows us to gauge audiences' responses to writers co-writing with AI and their perception of the authenticity of such work.

%Findings of the current study provide new insights into this knowledge space.
Our findings provide new insights into how writers and readers perceive co-writing with AI writing support. 
To begin with, writer-centered conceptions of authenticity focus more heavily on the internal experiences of writers and extend beyond the constructs of authenticity from prior literature (i.e., authenticity as source, category, and value of creators' work)~\cite{Auth_Humanity_Guignon_2008, Auth_Humanity_Handler_1986, Auth_Humanity_Art_Newman_Bloom_2012}.
Furthermore, the use of AI raises several broader questions for creative writing: \revision{Can writers still be regarded as the sole sources creating the content when AI writing tools are used? Can the resulting work still compellingly capture and speak for the writers' life experiences and the human stories informing the work?
%Can they still represent the (sole) sources of content creation when AI also contributes to the text? Can they still speak for the life experiences and human stories behind their work when AI helps organize writing materials into text? 
As writers reflected on these questions and practiced co-writing with AI during our study, they saw possible influences of AI-assisted writing on authenticity. While co-writing with AI did not fundamentally change their definitions (and thus understanding) of authenticity in creative work, they saw the need for and took various approaches to preserving their authentic voices in writing. This suggests an opportunity for the design of AI writing assistance tools to play an important role in supporting this endeavor.} %questions to the integrity of content producer and the representativeness of the human writer, posing threats to authenticity in writing.
%Meanwhile, the actual practices of co-writing with AI prompt writers to re-think the meaning of authenticity.
%This suggests that the design of AI writing assistance tools can play a key role in (re)shaping conceptualizations of authentic writing.
Finally, in contrast to writers' concerns, in our study, readers expressed great interest in \revision{reading AI-assisted writing and were curious about how AI's contributions might come into play.}

Our work makes three key contributions:
(1) We deepen our theoretical understanding of authenticity in the context of human-AI co-creation by surfacing and identifying writer-centered definitions of authenticity. These definitions incorporate aspects of authenticity that have not been accounted for in existing theories. %by highlighting both writers' and readers' points of view and extending theoretical constructs of authenticity in co-writing with AI.
(2) Our study offers design implications that \revision{underscore the need for co-writing tools, beyond their uses as productivity and creativity aids, to preserve authenticity in writing,} ranging from supporting writers' motivations and needs to addressing current pain points.
%such as adjusting to the formats and temporal considerations of AI writing assistance. 
%and to adjusting the formats and temporal considerations of AI writing assistance.
%Beyond possible aids for productivity and creativity--- often addressed in prior literature---our study's design implications place an additional requirement for preserving authenticity in writing.
%(3) To gauge the impact of generative AI at societal scales, we discuss our theoretical and design implications alongside the increasingly prevalent practices of personalizing AI applications, 
\revision{(3) Finally, we also draw broader implications for the design of personalized AI tools, as they have been adopted in more and more creative domains. We discuss whether individual creators' voices are amplified or lost when their own data is leveraged to create personalized AI tools.} %re-examining whether individuals' authentic voices are amplified or lost as they convey messages through AI-powered tools to their broader audiences.

\section{Background \& Related Work}

\subsection{Frameworks of Authenticity in Human Creative Work}
The humanistic literature has long acknowledged authenticity as the core of human creative work \cite{Auth_Humanity_Guignon_2008, Auth_Humanity_AMA_Lehman_2019, Auth_Humanity_Art_Newman_Bloom_2012}.
Though there is a longstanding history of proposing theoretical frameworks for authenticity, researchers have not yet formed a consensus on the definition of authenticity.
Still, they have often proposed three key themes to help define and conceptualize authenticity: \textit{Category}, \textit{Source}, and \textit{Value} \cite{Auth_Humanity_Guignon_2008, Auth_Humanity_Handler_1986, Auth_Humanity_AMA_Lehman_2019, Auth_Humanity_Creative_Persona, Auth_Humanity_Images_Charmé_1991, Auth_Humanity_Varga_Guignon_2023, Auth_Humanity_Art_Newman_Bloom_2012, Auth_Humanity_Lightness_Safran_2017, Auth_Humanity_Integral_Humanity_Anderson_1993, Auth_Humanity_Expressive_Authenticity_Lindholm_2013}.
First, the \textbf{\textit{Category}} theme concerns whether a piece of work matches one’s existing beliefs about its associated category. 
The scope of a category can vary, ranging from a particular style or school of work (e.g., Bauhaus-style architecture) to a certain era (e.g., a Renaissance painting).
Second, the idea of authentic \textbf{\textit{Source}} concerns whether one can trace a piece of work to a specific source (i.e., a person, a place, an event, or any type of origin).
This explains the importance of crediting writers and labeling the origins of work.
In particular, when work from certain individuals is truly one-of-a-kind---such as the highly recognizable work of Picasso---audiences can easily identify them as the source of content.
In such cases, the concepts of \textit{Category} and \textit{Source} become more blended and interchangeable. 
%Finally, \textit{\textbf{value and alignment}} addresses the consistency between a creator’s internal states and their external expression.
Finally, the \textit{\textbf{Value}} theme is about whether there is consistency between a creator’s internal states and their external expression.
\revision{This focuses on whether a writer's perspectives, opinions, and values are consistent with what they expressed in their work.}

In our study, we explore whether writers' perceptions of authenticity in writing align with these concepts of authenticity for broader creative work and whether specific concerns apply to writing. Based on writer-centered definitions of authenticity we uncover, we further examine whether co-writing with AI is considered authentic and whether emerging AI technologies change writers' views about the essence of authenticity.

%\textit{RQ1a: How do writers define authenticity in writing?}

%\textit{RQ1b: Based on their definitions of authenticity, do writers consider co-writing with AI authentic writing?}

%\textit{RQ1c: Whether and how do writers' responses to RQ1a and RQ1b change after co-writing with AI?}

\subsection{The Impact of AI Use on Authenticity in Writing}

\subsubsection{Writers' growing concerns regarding AI writing assistance}
The increasing popularity of using AI for creative tasks has motivated recent work to investigate the possible impact of AI on several aspects of creators' work, including credit, authorship, ownership, control, and agency~\cite{HAIC_authenticity_AI_Ghostwriter_Effect, HAIC_authenticity_Diffusing_Creator, HAIC_authenticity_Designing_Participatory_AI, HAIC_authenticity_choice_over_control}, many of which are closely related to authenticity.
Recent studies, workshops, panels, and other forms of discussion~\cite{news_Lee_2023} have thus far revealed mixed opinions from research communities, creators, and the general public toward these topics.
Here, we summarize a few emergent themes: %yet inconclusive themes:

\textbf{\textit{Writers remain hesitant to declare AI co-authorship publicly.}}
% [AI ghostwriter effect] 
%Writers don’t consider themselves as owners of AI-generated text, but they also refrain from publicly declaring AI authorship.
Recent work on perceived authorship of AI-generated text reveals a complex \textit{AI Ghostwriter Effect}~\cite{HAIC_authenticity_AI_Ghostwriter_Effect}.
Through comparing personalized and pseudo-personalized AI tools, the findings suggest that though writers did not perceive complete ownership over AI-generated text, they were also reluctant to publicly declare AI co-authorship.
This reluctance could be related to writers' concerns about negative responses from their readers~\cite{HAIC_authenticity_Social_Dynamics_Gero}.
But more importantly, writers feel that authorship should go hand in hand with the degree of contribution one makes to a piece of work~\cite{HAIC_authenticity_Diffusing_Creator}. %, although it remains unclear what constitutes \textit{contribution} from AI. 
% On the other hand, what makes meaningful human contributions when co-writing with AI may also remain an open question.
%Moreover, it is generally challenging to specify which parts of the writing process make more meaningful contributions than others.
However, it remains unclear what constitutes \textit{contribution} from AI, and it is generally challenging to specify which parts of the writing process make more meaningful contributions than others.

\textbf{\textit{Writers worry that using AI might negatively impact their writing outcomes.}}
In several studies, writers expressed concerns that AI might distract them from their original ways of writing, leading to lower quality of work~\cite{HAIC_authenticity_Designing_Participatory_AI}.
Writers often also raised the question of whether readers would be able to tell if they were using AI to write~\cite{HAIC_authenticity_Social_Dynamics_Gero}, and \revision{multiple studies suggest that audiences might react negatively if they knew a piece of text was generated by AI}~\cite{HAIC_authenticity_LLM_persuasion, HAIC_authenticity_Opinionated_LLM, HAIC_authenticity_Synthetic_Lies, AIwriting_self-presentation}.
It remains, however, unclear whether adverse reactions result regardless of the context in which AI is being used, or whether certain use cases of AI writing assistance are more acceptable.

\textbf{\textit{Writers worry that working with AI might diminish their control and joy during the writing process.}}
In general, writers hope to maintain control throughout their writing process~\cite{HAIC_authenticity_choice_over_control}.
Writers' sense of control shapes their perceptions of work ownership, which might in turn determine whether they are using AI as a tool or being influenced by AI~\cite{HAIC_tool_agency}.
As such, participants of a recent study expressed strong preferences for taking AI suggestions from multiple options instead of adopting a single, complete piece of AI-generated writing \cite{HAIC_authenticity_choice_over_control}.
From a job satisfaction perspective, writers also wondered whether working with AI would diminish the degree to which they enjoy the process of writing~\cite{HAIC_authenticity_Artificial_Colleague, HAIC_authenticity_Social_Dynamics_Gero}.

% [Social dynamics of AI support in writing] 
% Authenticity is a key value for writers. They are not only worried that (a) writing with AI would shift them away from their genuine voices but more about (b) whether their audiences would be able to identify such shifts in voices.
%  Related to (a): Prior work in AI-mediated communication suggests AI suggestions can shift writing tones, styles, and topics.
%  Related to (b): Prior work in writer-reader relationships suggests readers have the expectation for writers to put in their own work, and writers have the desire for genuine expressions.

% [Control over AI writing assistance] 
% Writers’ perceived control in the writing process shapes their perceived ownership of work.
% Meanwhile, agency of AI co-creation applications affects whether users perceive AI as tools (and have control over them) vs. as collaborative partners.

% Highlight literature gaps
Together, these emerging themes suggest that writers have growing concerns about AI's potential threats to authentic writing.
However, \textit{\textbf{it remains unclear whether and how writers' authentic voices can be preserved when writing with AI tools.}}
%Besides, these prior studies mostly focus on relevant yet specific aspects of authenticity only.
Prior work also tends to focus only on specific aspects of authenticity.
For instance, prior work studying authorship specifically addresses the \textit{Source} aspect of authenticity, while work examining the impact of AI assistance on writing outcomes tends to conceptualize authenticity more similarly to the \textit{Category} approach.
As a result, we lack a more holistic, writer-centered view of authenticity in human-AI co-writing.
Without comprehensive perspectives from writers, it remains difficult to propose solutions to regain and protect authenticity in their co-created work with AI that responds effectively to their needs and desires.
Our present research aims to address these literature gaps and design goals.

% \placeholder{Besides unresolved questions stemming from these studies, prior literature tends to focus on specific aspects that are partially relevant to authenticity. For instance, studying authorship specifically addresses the \textit{source} component of authenticity, while examining the impact of AI assistance on writing outcomes resonates with the \textit{category} approach. However, there lacks a more holistic view toward authenticity and the meaning of authentic work in human-AI co-creation. Among many of the above-mentioned studies, concerns for authenticity came out from writers, but research directly examining the subject of authenticity is scarce. Finally, despite growing concerns around this topic, \textbf{it remains unclear whether and how writers want to preserve authenticity in human-AI co-creation.} Without a clearer understanding of what authenticity is, it remains difficult to act toward addressing these concerns.}

\subsubsection{Insights from AI-mediated communication research}
Prior research on AI-mediated communication (AIMC) might bring additional perspectives on these unresolved questions around authenticity. In particular, AIMC studies have looked at how AI suggestions mediate written content for communication through text, including readers' perceptions of both the content and the message writer~\cite{AIMC_Hancock_JCMC_2020, AIMC_positivity_social_attraction, AIMC_topic_choice_self_presentation, AIMC_social_relationship, AIMC-perceptions-of-generated-problematic-replies}.
Importantly, writing can be viewed as a medium of communication that connects writers and their audiences, albeit often not bi-directionally or at the inter-personal level~\cite{writer-reader_Candlin_Hyland_1999, writer-reader_sharing_words, writer-reader_writing_for_readers}.

% evidence that AI assistance could mediate content output --> shift authentic voices in writing
Indeed, various AIMC studies have shown that AI assistance (e.g., word-by-word suggestions or short-reply suggestions) can shift the characteristics of writing content, such as a more positive tone~\cite{AIMC_positivity_social_attraction, AIMC_Hancock_JCMC_2020, AIMC-perceptions-of-generated-problematic-replies}, or even affect topics addressed in writing~\cite{AIMC_topic_choice_self_presentation}.
More recently, \cite{HAIC_authenticity_Opinionated_LLM} found that writing with AI assistance on opinionated content affects not only a user's writing output but also their own opinions toward the written topic.
In light of these findings as well as other repeated findings around readers' negative perceptions of communicators who use AI~\cite{jakesch2019ai,liu2022will}, scholars have expressed ethical concerns around AI's implications for information and interpersonal trust~\cite{AIMC_email_trust, AIMC_uncertainty_source}, users' sense of agency and authenticity~\cite{AIMC-perceptions-of-generated-problematic-replies}, and social relationships~\cite{AIMC_social_relationship, AIMC_positivity_social_attraction, AIMC_profile_airbnb} as users adopt AI more widely for composing content for interpersonal communication.

While these findings suggest AI's potential to move writers away from their own voices in writing,
% Furthermore, AI mediation may affect writers' opinions and attitudes during their writing processes \tocite{},
it is not clear whether insights from AIMC research apply to understanding AI-assisted writing more generally beyond interpersonal communication.
Once again, we know little about how writers might avoid compromising their own voices when writing with AI.
In particular, the connections between professional writers and their readers through writing are often different than the forms of interpersonal communication studied in AIMC, which often examines content written in mutual conversations for instrumental purposes (e.g., communicating with an owner of a rental house) or relationship-building (e.g., emailing a friend)~\cite{AIMC_profile_airbnb, AIMC_Hancock_JCMC_2020,liu2022will}.
Unlike such dialogues, writers seldom speak to a particular individual or aim to build interpersonal relationships with their audiences.
Instead, writers use writing as a means to deliver their messages to a crowd and express themselves creatively.
We thus investigate how these differences add to the complexity of authenticity in human-AI co-writing.

\subsection{Designing Tools for Human-AI Co-Writing}

\begin{table}[b]
    \centering
    \resizebox{\textwidth}{!}{%
    \begin{tabular}{p{0.08\textwidth} | p{0.14\textwidth} | p{0.14\textwidth} | p{0.25\textwidth} | p{0.25\textwidth} | p{0.35\textwidth}}
    \toprule
        Literature & Use of language model & Types of writing & Forms of AI writing assistance & Key variables, conditions, and/or prototypes examined & Key findings \\
        \hline
        \cite{AIwriting_opinionated_LLM} & GPT-3 & Argumentative writing & Text continuation/completion through word-by-word suggestions & AI suggestions prompted with positive/negative opinions of a given topic & Co-writing with an opinionated AI writing assistance shifted opinions expressed in users' writing and their attitudes toward the written subsequently. \\
        \hline
        \cite{AIwriting_self-presentation} & GPT-2 & Self-introduction in online profiles & Text continuation/completion through word-by-word suggestions & AI suggestions fine-tuned with specific topics & Adopting AI suggestions altered what users wrote about for self-presentation. \\
        \hline
        \cite{AIwriting_multiword_non-native} & GPT-2 & Business writing in emails & Text continuation/completion through multi-word suggestions & Length of AI suggestions & (1) Multi-word suggestions benefited ideation at the cost of efficiency in writing; (2) Non-native speakers benefited more from multi-word suggestions. \\
        \hline
        \cite{AIwriting_BunCho_Japanese} & GPT-2 & Short text of creative writing & Full-text generation & An interactive interface that allowed users to prompt AI writing assistance with topical and atmospheric keywords & Writers enjoyed writing with AI more than conventional writing processes and suggested AI output broadened the topics they would write about. \\
        \hline
        \cite{AIwriting_creative_writing_RNN} & an RNN-based language model & Creative writing & Text continuation/completion through sentence-level suggestion & Degree of randomness (temperature) in AI suggestions & AI suggestions with lower randomness were more coherent with users' own writing, while increased randomness introduced more novel suggestions. \\
        \hline
        \cite{AIwriting_sparks_science-writing_Gero} & GPT-2 & Science writing & Generating "sparks" (sentences related to a given scientific concept) & AI-generated topical sentences as inspiration for writers & (1) AI-generated "sparks" provided inspiration, translation, and perspectives to writers; (2) The quality of "sparks" did not affect users' satisfaction of writing assistance. \\
        \hline
        \cite{AIwriting_metaphorian} & GPT-3 & Science writing & Generating metaphors of given scientific topics & AI-generated metaphors as inspiration for writers & AI-generated metaphors offered inspiration and created positive user experiences without compromising writers' agency. \\
        \hline
        \cite{AIwriting_Storyfier_vocabulary_learning} & Hugging Face T5 & Writing as a practice to support vocabulary learning & Generating stories based on given keywords & AI-generated stories as learning support & Users favored learning vocabularies with AI-generated stories but yielded worse learning outcomes. \\
        \hline
        \cite{AIwriting_visual_draft} & GPT-3.5-turbo & Argumentative writing & Visualizing structures of written content to support writers to prototype their drafts & Visual prototyping with drafts & Visual prototyping allowed writers to experimented with and validated their ideas in drafts. \\
        \hline
        \cite{AIwriting_auto_summary} & Hugging Face T5 & Analytic essay & Text summarization & Real-time summarization of users' writing & Summarization helps users gain insights and strategize their own work. \\
        \hline
        \cite{AIwriting_Tool_arbitrary_style_transfer} & LaMDA & Rewriting text into a given style & Style transfer & Effectiveness of style transfer through zero-shot learning and framing style transfer as a sentence rewriting task & The method (augmented zero-shot learning) proposed in the present work can accomplish standard style transfer tasks (e.g., changing the sentiment of text) and also be used for re-writing text metaphorically. \\
        \hline
        \cite{AIwriting_Tool_wordcraft, AIwriting_tool_wordcraft_creative_writing} & LaMDA & Creative writing & Idea suggestion, text continuation, expansion, rephrasing, and re-writing & An co-writing interface with multiple forms of AI writing support & Designing writing support for experts need specific considerations as they have more developed writing practices and styles. \\
        \hline
        \cite{AIwriting_Tool_integrative_leap_multimodal} & GPT-2 & Creative writing & Text + image generation per users' prompt input & Benefit of providing multimodal output as sources inspiration for writers & While multimodal output could be inspiring, it also took writers more work to integrate them into their current work. \\
        \hline
        \cite{AIwriting_tool_screenplay} & Chinchilla AI & Screenplay & Dialogue generation based on story world-building specified by writers & Effectiveness of prompting with high-level narrative arc to generate dialogues for screenplay & Writers could better focus on the overarching story building when working with this tool. \\
    \bottomrule    
    \end{tabular}
    }
    \caption{Recent literature on human-AI collaborative writing. Here, we only include studies that examine how human users co-write with AI and the outcomes of co-writing. As such, work that studies the model/system design of AI writing assistance \textit{without} human evaluation is beyond the scope of our literature review.}
    \label{tab:AI_writing}
\end{table}

Recent work has already proposed a wide variety of possible AI writing assistance scenarios \revision{(see our summary in Table~\ref{tab:AI_writing})}, ranging from more generic to genre-specific support. 
Researchers have also experimented with AI assistance that supports different stages of one's writing process as well as providing multi-form~\cite{AIwriting_Tool_wordcraft, AIwriting_tool_wordcraft_creative_writing} or multimodal~\cite{AIwriting_Tool_integrative_leap_multimodal} support at once.
While many of these explorations have been shown to be helpful for writers, the majority of these studies focus on improving usability, productivity, and sometimes creativity of writing. 
By contrast, the possible influences of these tools on authenticity often only arise in exploratory analyses or post-study conversations with participants, leaving a noticeable literature gap.

% Other papers on AI-assisted writing tool:
% https://dl.acm.org/doi/abs/10.1145/3490099.3511105
% https://dl.acm.org/doi/abs/10.1145/3544548.3581225
% https://dl.acm.org/doi/full/10.1145/3511599 (also proposed the concept of “integrative leaps” when working with AI)
% https://dl.acm.org/doi/abs/10.1145/3290605.3300526
% https://dl.acm.org/doi/abs/10.1145/3532106.3533533
% https://dl.acm.org/doi/abs/10.1145/3526113.3545672

% We explore this topic through personalized vs. non-personalized tools.
% \textit{RQ5: What types of support do writers seek in creative writing with AI?}

% \subsection{Personalized Support in Human-AI Co-creation}
% \textit{RQ6: Whether and how do personalizing co-writing experiences with AI affect authenticity in writing?}

\section{Methods}

\subsection{Overview of the Study Design}
\begin{figure}[h!]
    \centering
    \includegraphics[width=\textwidth]{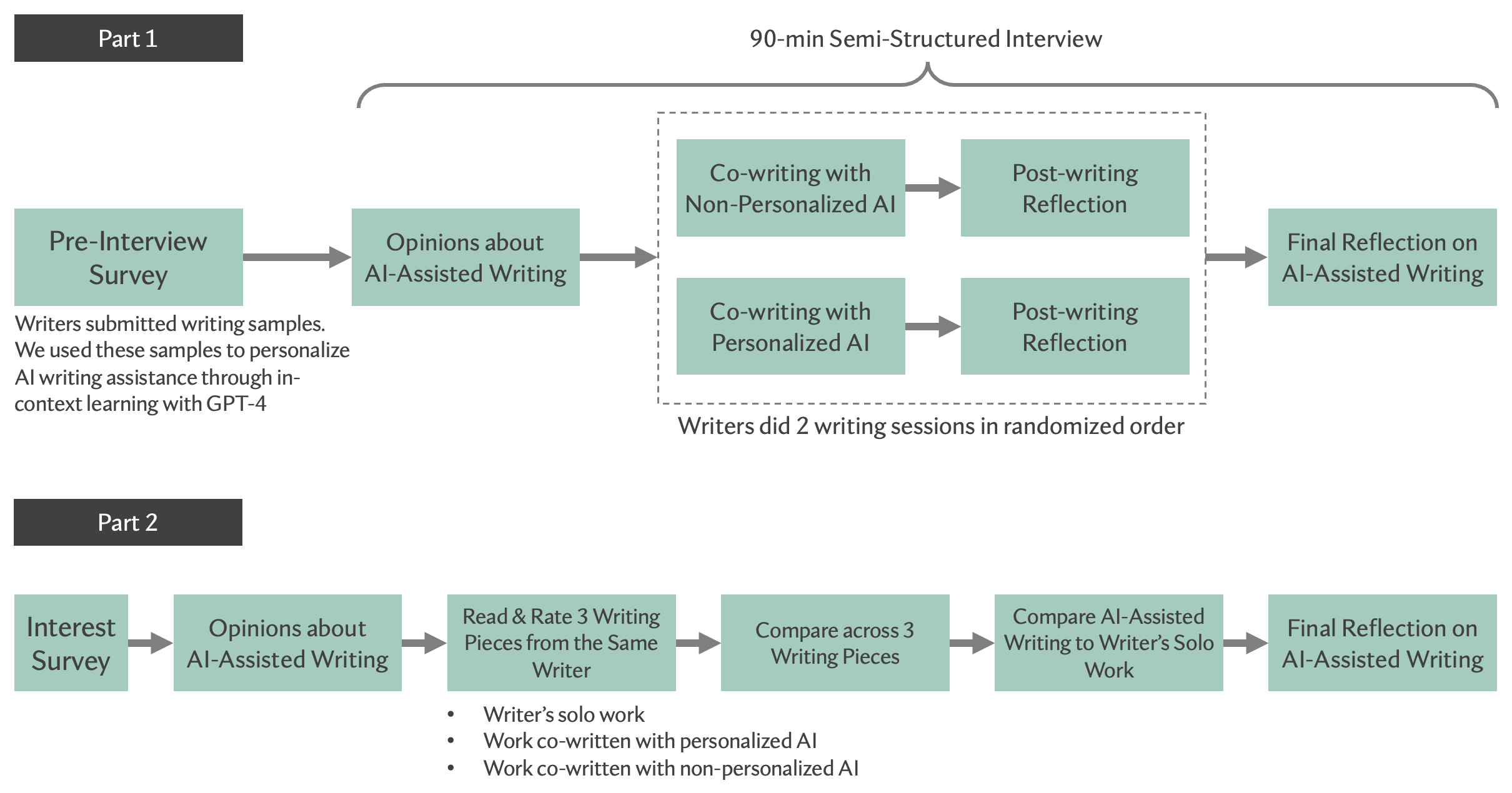}
    \caption{Study procedure for Part 1 (interview study with writers) and Part 2 (survey study with readers).}
    \label{fig:study_flow}
\end{figure}

We examined writer-centered definitions of authenticity and the impact of AI writing assistance on authentic writing primarily through semi-structured, qualitative interviews with professional writers. 
To answer RQ3 and to enable participants to respond to our inquiries with situated experiences, we adopted two versions of AI-powered writing assistance tools: one with personalization through in-context learning, and one without personalization. 
We investigated how participants wrote with these tools in real time and delved into their co-writing experiences through interviews following each writing session. 
We complemented perspectives from these direct users of such emerging technology (i.e., {\em writers}) through a follow-up online study with indirect stakeholders (i.e., {\em readers}) \cite{Friedman_value_sensitive_design}. 
Through these two parts of the study, we synthesize a more comprehensive view of authenticity in writing.
The full study protocol \revision{(as illustrated in Figure~\ref{fig:study_flow})} was reviewed and approved by the Institutional Review Board (IRB) of the authors' affiliation.

\subsection{Part 1: Interviews with Professional Writers}
We conducted Part 1 of the study with 19 professional writers.
The study includes a pre-interview survey (through an online questionnaire) where participants shared their writing samples and professional experiences as writers, and expressed their opinions on what authenticity in writing means to them and on AI writing assistance.
During each interview, we first asked participants to further elaborate on their perspectives regarding authenticity and co-writing with AI.
Next, participants engaged in two writing sessions to co-write a short passage with a personalized and a non-personalized AI writing assistance tool in a randomized order.
Immediately after each writing session, we asked participants to reflect on their co-writing experiences through semi-structured interviews.
After both writing sessions, we asked participants to compare their experiences across the two sessions.
Finally, we revealed to them the difference between the two tools and asked them a few more questions to capture their overall thoughts on co-writing with generative AI.
The pre-study survey and the full interview protocol are included in Appendix~\ref{appendix:pre-survey} and \ref{appendix:interview_protocol}.

% \todo{add a study flow chart}

\subsubsection{Recruitment and participants}
We recruited 21 professional writers through Upwork, a platform frequently used to recruit professionals with specific expertise. 
We partnered with a recruitment specialist from the Upwork team to ensure candidates provided authentic information on their profiles and verified their past freelance work on the platform.
We also used the pre-interview survey for screening purposes, in which participants elaborated on their professional experiences as writers, shared their writing samples, and provided relevant links to their professional websites and profiles (see more details of the pre-interview survey in Section~\ref{sec:writer_pre-survey}).
One of the participants did not show up and one decided to drop out from the study, resulting in N = 19 participants who fully completed the study; data from all 19 participants was used for analyses.
\revision{In Table~\ref{tab:writer}, we report the background and experiences of these writer participants.}

\revision{\textit{A note on sample size:} Like the majority of qualitative research, we did not have a definite approach to determining the exact sample size. However, we followed recommendations from prior methodological reviews~\cite{qual_sample_size_2013, qual_standard_2016} such that (1) we ensured the number of interviews conducted for the same task fell between $15-30$; (2) we referred to the sample sizes of qualitative studies in relevant publication venues (e.g.,~\cite{AIwriting_sparks_science-writing_Gero, HAIC_authenticity_Social_Dynamics_Gero, AIwriting_Tool_wordcraft}); and (3) throughout the study period, we observed that emergent themes arose and saturated even accounting for new interviews. Despite the relatively small sample size, we ensured that participants had diverse backgrounds to improve the generalizability of our results, such that our findings were not confounded with participants' seniority, experience, and writing genre.}
%Our primary strategy was to recruit writers of diverse backgrounds, such that our findings were not confounded with their seniority, experiences, and the types of work produced.}

\begin{table}[b]
    \centering
    \renewcommand{\arraystretch}{1.2} 
    \resizebox{0.96\textwidth}{!}{%
    \begin{tabular}{p{0.1\textwidth}|p{0.2\textwidth}|p{0.12\textwidth}|p{0.55\textwidth}}
    \toprule
        \textbf{Participant} & \textbf{Primary\newline Writing Genre} & \textbf{Years of\newline Experience} & \textbf{\textcolor{black}{Experience with genAI}}  \\ % topics of writing samples?
        \hline
        P1 & Science fiction, fantasy &  5 - 10 & \textcolor{black}{P1 regularly used genAI tools but only for text editing purposes.} \\
        \hline
        P2 & Comedy, science fiction & 10 - 15 & \textcolor{black}{P2 frequently used genAI tools for a wide range of purposes and was familiar with genAI models, such as GPT-3 and GPT-4.}\\
        \hline
        P3 & Lifestyle & $<$5 & \textcolor{black}{P3 regularly used genAI tools for text editing purposes (specifically for personal blog posts). They occasionally used genAI to brainstorm ideas for titles of their written work.}\\
        \hline
        P4 & Science writing & $<$5 & \textcolor{black}{P4 regularly used genAI tools but only for text editing purposes.}\\
        \hline
        P5 & Fantasy & $<$5 & \textcolor{black}{P5 had no experience with genAI at the time of the study.} \\
        \hline
        P6 & Lifestyle & 5 - 10 & \textcolor{black}{P6 regularly used genAI tools but only for text editing purposes.}\\
        \hline
        P7 & History & 15 - 20 & \textcolor{black}{P7 regularly used genAI tools to rewrite and make their drafts clearer.}\\
        \hline
        P8 & Science fiction & 10 - 15 & \textcolor{black}{P8 had no experience with genAI at the time of the study.}\\
        \hline
        P9 & Lifestyle & $<$5 & \textcolor{black}{P9 regularly used genAI tools but only for text editing purposes.} \\
        \hline
        P10 & Biography & $>$20 & \textcolor{black}{P10 had no experience with genAI at the time of the study.}\\
        \hline
        P11 & Science writing & 5 - 10 & \textcolor{black}{P11 had experimented with and explored the capabilities of various genAI tools. They did not find these tools useful for their writing jobs and thus did not use them regularly.}\\
        \hline
        P12 & Poetry & $>$20 & \textcolor{black}{P12 regularly used genAI tools to conduct research for their writing and used them for text editing purposes.} \\
        \hline
        P13 & Horror & 10 - 15 & \textcolor{black}{P13 occasionally used genAI tools to write informational content.} \\
        \hline
        P14 & Romance & 5 - 10 & \textcolor{black}{P14 used genAI tools for general purposes but not for writing.} \\
        \hline
        P15 & Spiritual & 5 - 10 & \textcolor{black}{P15 had no experience with genAI at the time of the study.} \\
        \hline
        P16 & Romance & 5 - 10 & \textcolor{black}{P16 regularly used genAI tools but only for text editing purposes.}\\
        \hline
        P17 & Science fiction & $<$5 & \textcolor{black}{P17 regularly used genAI tools but for text editing purposes; such practice was requested by many of their clients. For a specific client, the participant was provided with informational content generated by genAI and was asked to provide "human voice" to the writing.}\\
        \hline
        P18 & Fantasy & 5 - 10 & \textcolor{black}{P18 frequently used genAI tools to generate written content for a client.}\\
        \hline
        P19 & Screenplay & $>$20 & \textcolor{black}{P19 regularly used genAI tools to generate visual content accompanying their writing work. They also frequently used genAI tools to help with formatting their writing.} \\
        \bottomrule
    \end{tabular}}
    \caption{Professional profiles of writer participants in Part~1}
    \label{tab:writer}
\end{table}

\subsubsection{Pre-interview survey}
\label{sec:writer_pre-survey}
The pre-interview survey begins with obtaining informed consent from participants (by signing a consent form).
After consenting, participants responded to a series of questions through free text responses.
These questions ask participants to describe (1) the unique characteristics of their writing, (2) their prior experience working with AI and non-AI writing support tools, if any, (3) their definition of authenticity in writing.
At the end of the survey, participants were asked to submit a short writing sample of their own (~ 200 words) that best represents their writing. Participants were informed their writing samples would be used as to design the study task for them (if selected to participate), though we did not specify how so.
These writing samples were used for in-context learning to prepare a personalized AI writing assistance tool for the interview session (see more details in the next section), and a subset were used in the online survey with readers (see Part 2 Method in Section~\ref{sec:part2_method}).

\subsubsection{AI writing assistance}
\begin{figure}
    \centering
    \includegraphics[width = 0.6\textwidth]{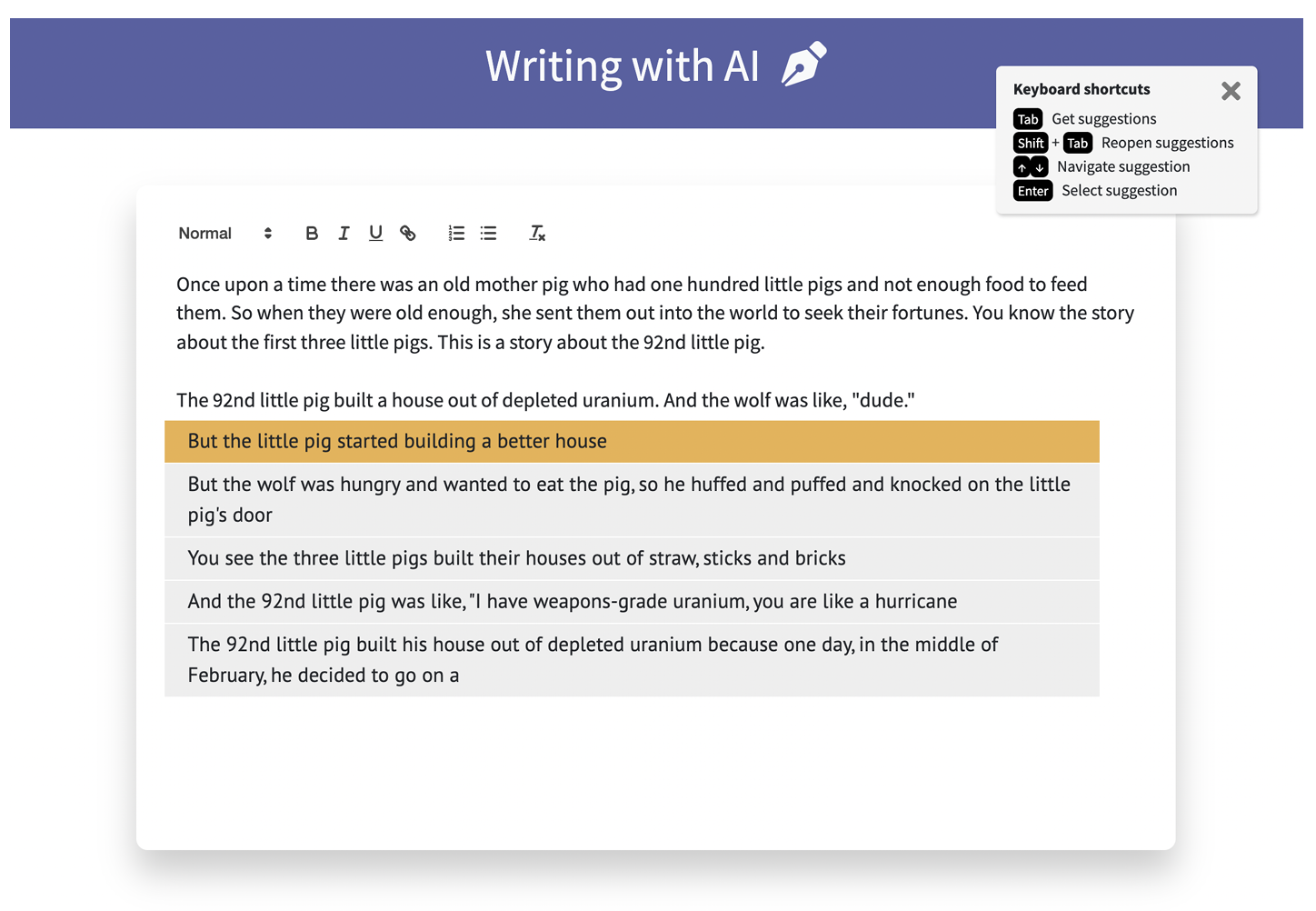}
    \caption{Interface of the CoAuthor system (image adopted from \cite{Mina_Lee_CoAuthor}), which contains a text editor where users can request (Tab key) and adopt AI (Enter key) suggestions through keyboard shortcuts.}
    \label{fig:coauthor}
\end{figure}

We adopted open-source code from \cite{Mina_Lee_CoAuthor} to run the CoAuthor interface (see Figure~\ref{fig:coauthor}) with GPT-4 and used it as the AI writing assistance tool for the interview sessions. CoAuthor is a system built by Lee et al. \cite{Mina_Lee_CoAuthor} that allows writers to request next-sentence suggestions from an LLM, and the open-source code allows for customization, such as choosing which LLM to use and its parameters. % \VL{the paper never includes an interface...should we add?}
We chose CoAuthor as the writing assistance tool for this study for two reasons: 
(1) As we targeted writers with expertise in a diverse set of writing genres, we chose a tool that was \textit{not} designed to support a particular type of writing. 
(2) Since participants had a limited time to become familiar with the AI writing assistance, we avoided more complex tools with multiple forms of support (e.g., Wordcraft~\cite{AIwriting_Tool_wordcraft}).

%(3) Though CoAuthor only supports a single type of writing assistance (i.e., sentence-level suggestions), it goes beyond word-by-word completion so that we can better differentiate the findings of the present work from those revealed in prior AIMC research. 
%\VL{AIMC works also studied sentence suggestion, and if you mention this, should we attribute our results difference just to the sentence level suggestions? I worry mentioning this might hurt the paper}
%\angel{right, this last point could be misleading - commenting it out now}

Before each session, we used the writer's submitted writing sample as training data for in-context learning to create a personalized version of the CoAuthor tool for their session.
We used the following parameters for GPT-4 to generate suggestions: max tokens = 5; temperature = 0.9; top p = 1; presence penalty = 0.5; frequency penalty = 0.5.
During the co-writing sessions, writers were able to request AI writing suggestions at any time by the Tab key.
The tool would generate around 5 sentence-level suggestions based on the written text.
Writers could accept, reject, and revise the suggestions as they liked.
At the beginning of their first writing session, the researcher who led the study session walked the writers through the CoAuthor interface and guided them to practice using the tool to write a few sentences. 
This was to ensure participants fully understood and became familiar with the tool before they started their first formal writing session.

\subsubsection{Co-writing session with AI}

Participants were asked to write one short passage ($\sim$200 words) for each co-writing session; participants chose to write on a topic similar (but not identical) to that of their submitted writing sample. %\todo{add writing topic to participant table}
Each co-writing session lasted between 20 - 25 minutes. The time was determined by piloting with two professional writers.
All participants were able to finish their assigned 200-word passage within the time frame.
As participants wrote, the CoAuthor interface recorded their writing logs and their final writing output.
Writers' behavioral data recorded through the writing logs include (1) the frequency and timing of their AI suggestion requests, (2) the frequency and timing of their acceptances and/or rejections of AI suggestions, (3) the AI text suggestions provided at each of their requests, (4) the AI suggestions (in text) accepted, if any, and (5) the text inserted by writers. 
Additional details about the writing logs can be found in Appendix~\ref{appendix:writing_log}.
After each writing session, participants participated in a short interview ($\sim$15 minutes) to reflect on their writing experiences.

\subsection{Part 2: Online Surveys with Avid Readers}
\label{sec:part2_method}
In Part~2, we conducted an online study to understand avid readers' perceptions of authenticity, AI writing assistance, and work co-written by human writers and AI.
We used writing work produced by six writers in Part~1, each of which represents a unique literature genre (P8: science fiction/fantasy, P10: biography, P12: poetry, P14: spiritual, P15: romance, and P19: screenplay), as materials for Part~2.\footnote{At the end of each session in Part 1, we informed writers and obtained their consent to use their writing for subsequent research.}
These include anonymized writing samples that writers submitted in the pre-interview survey of Part~1 (i.e., work written by themselves) as well as the two passages they wrote with the personalized and non-personalized AI during Part~1.

Participants of Part 2 (i.e., readers) first responded to a few open-ended questions about their attitudes toward AI-assisted writing.
They then read three writing passages from the same writer (their solo work, work with personalized AI, and work with \revision{non-}personalized AI) in randomized order.
After reaching each article, participants assessed the writing with respect to \textit{likeability} (how much they liked the writing), \textit{enjoyment} (how much they enjoyed reading the writing), and \textit{creativity} (how creative they thought the writing was).
We chose these evaluation criteria as they were also used in prior work to assess AI-assisted writing (e.g.,~\cite{AIMC_profile_airbnb, AIMC_email_trust}).

Next, readers performed two sets of comparisons with the three pieces of writing.
First, they were informed that \textit{some} of these pieces were written with AI, and were asked to compare and rate how likely each piece was to have been written by a human writer alone, or with the help of an AI writing support tool.
%Readers were informed the three pieces of writing were either written independently by a human writer or co-written by the same human writer and an AI writing assistance tool. 
Second, readers were informed which piece was the writer's solo work.
They then compared the two AI-assisted writings \revision{by responding to the question: \textit{``Compared to the text written independently by the author, to what extent do you think the co-written text preserves the authentic voice of the author?''}}
%, rating how much they sound similar to or different from the writer's voice in the solo writing. 
They were also asked to identify in which part(s) of the writing the author might have adopted AI assistance.
Readers were blinded to the two different AI conditions; namely, they did not know which passage was co-written with a personalized versus a non-personalized AI tool, or that personalization was the variable being studied.
We applied this blinding approach to ensure participants' assessments were based on the content per se and were not affected by their perceptions of personalized AI assistance.
%They also did not know personalization was the manipulated variable in Part~1.

After rating and comparing writing passages, participants responded to a few questions about their general opinions about AI-assisted writing.
Table~\ref{tab:part2_procedure_measure} presents the procedure of Part~2 and the variables measured throughout.
Additionally, the full Part~2 questionnaire is attached in Appendix~\ref{appendix:part2_survey} and \revision{descriptive statistics of numeric measures are reported in Table~\ref{tab:descriptive}}.
The entire Part~2 study was conducted online and took around 45 - 60 minutes to complete.
Participants received cash compensation for their participation.
Once again, the study protocol was reviewed and approved by the IRB of the authors' affiliation.

\begin{table}
\fbox{%
\parbox{0.95\linewidth}{

\textbf{Step 1.) At the beginning of the study, participants responded to a few questions to express their opinions about AI writing assistance.} These are single measures rated with 5-point Likert scales, including:
\begin{itemize}
    \item \textit{Familiarity with generative AI:} Asking participants what they know about generative AI.
    \item \textit{Interest in reading AI-assisted writing:} Asking participants whether they are interested in reading work co-written with AI.
    \item \textit{Authenticity of AI-assisted writing:} Asking participants whether they consider AI-assisted writing as a writer's authentic work.
\end{itemize}
\vspace{0.15cm}

\textbf{Step 2.) Participants read three pieces of work from the same writer, including their solo work, work co-written with personalized AI, and work co-written with non-personalized AI.} Participants were blinded from the three conditions. That is, they did not know which piece was done by the writers independently and which with AI assistance.
\vspace{0.2cm}

\textbf{Step 3.) Participants responded to a few questions to express their perceptions toward the three pieces of work while blinded from the three conditions.} That is, participants did not know which piece was done by the writers independently and which with AI assistance when they responded to these questions. These are repeated measures; i.e., the same scales are used to rate each of the three pieces of work.
\begin{itemize}
    \item \textit{Likeability:} Asking participants how much they like a piece of writing. 
    \item \textit{Enjoyment:} Asking participants how much they enjoy a piece of writing.
    \item \textit{Creativity:} Asking participants how creative a piece of writing is.
\end{itemize}
\vspace{0.15cm}

\textbf{Step 4.) Participants were told some of these writings could have been done with AI assistance. They responded to a few questions to gauge which piece might be the writer's solo work and which piece might be co-written with AI.} These are repeated measures; i.e., the same scales are used to rate each of the three pieces of work.
\begin{itemize}
    \item \textit{Likelihood of human writing:} Asking participants to gauge whether a piece of writing was written by a writer independently or was co-written with AI assistance.
\end{itemize}
\vspace{0.15cm}

\textbf{Step 5.) Participants were told which piece of work was the writer's solo work and which two were co-written with AI, though we did not reveal which was done with personalized AI and which was done with non-personalized AI. Participants responded to a few questions to compare the two AI-assisted pieces to the writers' solo work.} These are repeated measures; i.e., the same scales are used to rate the two pieces of AI-assisted work.
\begin{itemize}
    \item \textit{Preserving writer's authentic voices:} Asking participants whether a piece of AI-assisted writing preserves a writer's authentic voice.
    \item Credits and authorship of work: Asking participants how they might attribute the credits and authorship of a piece of AI-assisted work.
\end{itemize}
\vspace{0.15cm}

\textbf{Step 6.) Participants responded to a few questions to reflect on their opinions toward AI-assisted writing after reading some samples.} These are single measures rated with 5-point Likert scales, including:
\begin{itemize}
    \item Perception of the writing: Asking participants' overall perception toward AI-assisted writing.
    \item Perception of the human writer: Asking participants' perception toward writers who apply AI assistance to complete their work.
    \item Appreciation and evaluation of writing: Asking participants whether they would appreciate a piece of AI-assisted work through different approaches.
\end{itemize}
}
}
%\vspace{0.1cm}
\caption{The procedure and variables measured throughout Part~2 of the study.}
\label{tab:part2_procedure_measure}
%\vspace{-0.3cm}
\end{table}

\begin{table}[t]
    \centering
    \resizebox{0.99\textwidth}{!}{
    \begin{tabular}{p{0.4\textwidth}|p{0.2\textwidth}|p{0.2\textwidth}|p{0.2\textwidth}}
        \toprule
        Variable & Round 1 & Round 1 (removed 7 possibly AI-generated responses) & Round 2 \\
        \hline
        \rowcolor{lightgray!20!}
        \multicolumn{4}{c}{\textbf{Opinions about AI writing assistance}} \\
        \hline
        \textbf{Familiarity with generative AI} & $3.85\pm0.92$ & $3.78\pm0.96$ & $3.65\pm1.00$\\
        \hline
        \textbf{Interest in reading AI-assisted writing} & $3.58\pm0.89$ & $3.76\pm0.91$ & $3.69\pm0.92$ \\
        \hline
        \textbf{Authenticity of AI-assisted writing} & $3.70\pm0.72$ & $3.30\pm0.98$ & $3.44\pm0.95$\\
        \hline
        \rowcolor{lightgray!20!}
        \multicolumn{4}{c}{\textbf{Reading writing samples}} \\
        \hline
        \textbf{Likeability} &    &   &   \\
        Writer's solo work & $3.80\pm0.77$ & $3.85\pm0.77$ & $3.88\pm0.77$\\
        Work co-written with personalized AI  & $3.55\pm0.51$ & $3.7\pm0.54$ & $3.65\pm0.54$ \\
        Work co-written with non-personalized AI  & $3.75\pm0.72$ & $3.85\pm0.72$ & $3.74\pm0.79$ \\
        \hline
        \textbf{Enjoyment} &    &   &   \\
        Writer's solo work &   $3.55\pm0.89$ & $3.63\pm0.88$ & $3.71\pm0.91$ \\
        Work co-written with personalized AI  & $3.40\pm0.88$ & $3.52\pm0.89$ & $3.53\pm0.86$ \\
        Work co-written with non-personalized AI  & $3.4\pm0.88$ & $3.59\pm0.97$ & $3.59\pm0.96$ \\
        \hline
        \textbf{Creativity} &    &   &   \\
        Writer's solo work & $3.70\pm0.98$ & $3.70\pm0.99$ &  $3.76\pm0.99$ \\
        Work co-written with personalized AI  & $3.70\pm0.80$ & $3.59\pm0.75$  & $3.59\pm0.70$ \\
        Work co-written with non-personalized AI  & $3.25\pm0.72$ & $3.48\pm0.80$ & $3.47\pm0.79$ \\
        \hline
        \rowcolor{lightgray!20!}
        \multicolumn{4}{c}{\textbf{Comparing all writing samples}} \\
        \hline
        \textbf{Likelihood of human writing} & & & \\
        Writer's solo work & $3.63\pm1.01$ & $3.73\pm0.92$  & $3.73\pm0.91$ \\
        Work co-written with personalized AI  & $2.84\pm1.01$ & $2.96\pm1.00$ & $3.00\pm1.00$ \\
        Work co-written with non-personalized AI  & $3.32\pm1.2$ & $3.19\pm1.13$ & $3.09\pm1.10$ \\
        \hline
        \rowcolor{lightgray!20!}
        \multicolumn{4}{c}{\textbf{Comparing personalized vs. non-personalized writing samples}} \\
        \hline
        \textbf{Preserving writers' authentic voices} &    &   &   \\
        Work co-written with personalized AI  & $3.20\pm0.68$ & $3.32\pm0.89$ & $3.30\pm0.82$ \\
        Work co-written with non-personalized AI  & $3.6\pm0.91$ & $3.36\pm0.90$ & $3.37\pm0.84$ \\
        \hline
        \textbf{Credits and authorship of work} &    &   &   \\
        Work co-written with personalized AI  & $2.90\pm0.79$ & $3.07\pm0.83$ & $3.09\pm0.79$ \\
        Work co-written with non-personalized AI  & $3.45\pm1.00$ & $3.30\pm0.99$ & $3.27\pm0.94$ \\
        \hline
        \rowcolor{lightgray!20!}
        \multicolumn{4}{c}{\textbf{Opnions after reading AI-assisted writing}} \\
        \hline
        \textbf{Perception of the writing} & $3.35\pm0.97$ & $3.30\pm0.98$ & $3.35\pm0.91$ \\
        \hline
        \textbf{Perception of the human writer} & $2.95\pm1.25$ & $3.48\pm1.11$ & $2.97\pm1.25$ \\
        \hline
        \textbf{Appreciation and evaluation of writing} & $3.50\pm1.21$ & $2.96\pm1.21$ & $3.56\pm1.04$  \\
        \bottomrule
    \end{tabular}}
    \caption{Descriptive statistics of numeric variables in Part~2. Numeric values in all cells represent $Means \pm S.D.$}
    \label{tab:descriptive}
\end{table}

\subsubsection{Recruitment and participants}
We recruited participants on Reddit to find %highly involved 
avid readers who regularly read the particular literature genres written by the six writers chosen from Part~1.
One of the researchers joined multiple subreddits dedicated to discussions about reading.\footnote{We received approvals from community moderators to post and participate in the discussions on these subreddits.} (See Appendix~\ref{appendix:subreddit} for a list of these subreddits.)
%Many of these Subreddits also focus on specific literature genres.
Recruitment messages were posted on discussion threads in these subreddits.
Interested individuals first filled out an interest form that surveyed the literature genres they were interested in and how frequently they read. They also reported the subreddit where they saw the recruitment message.
\revision{30.27\% of those who filled out our interest form read on a daily basis, and 60.54\% of them read weekly.}
We recruited participants who read their genres of interest at least once a week \revision{(Appendix~\ref{appendix:reader_genre} shows a list of genres that our reader participants regularly read. Note that each participant might frequently read more than one genre.)}.
As prior research on Reddit has identified a large number of machine-generated responses, we showed some question texts in image format in the interest form to filter out bots that can only read text.
We also included an open-ended question (i.e., ``In one to two sentences, describe your favorite literature genre and why you favor it.'').
Two researchers reviewed these open-text responses and screened out those that were likely generated by machines, as detailed below.

\subsubsection{Screening AI-generated responses}
We first set out to collect feedback from three readers for each writer's work, resulting in $6\times3 = 18$ participants.
However, we noticed a substantial number of open-text responses that might be AI-generated (e.g., by ChatGPT) after examining our initial batch of data.
We discussed among our research team to identify the following criteria for screening: (1) Responses that were consistently written in a bullet-point style following the structure of "[Topic]: [Description]," which is a common style in ChatGPT output; (2) We used our survey questions as prompts and collected responses from popular generative AI tools, including ChatGPT, Microsoft Bing Chat, and Google Bard. We cross-checked whether participants' responses overlapped significantly with those AI-generated ones.

Two researchers independently read through these responses and marked responses that were likely AI-generated. 
Any disagreement was reviewed by a third researcher.
In the end, we marked 7 participants as likely having used AI to generate open-text responses throughout the study and removed their data from analyses.
We thus conducted a second round of data collection and applied the same screening criteria to attain a larger sample size with sufficient power for data analysis. 
We targeted having at least 5 readers rating each passage; 
the final sample size after Round 2 of data collection is N = 30.
For all analyses, we reported results using data from both rounds of data collection.

\section{Results}

\subsection{Overview of Data Analysis}
We conducted analyses with three streams of data: a)~the interview data and b)~writing logs (i.e., behavioral data) from writers in Part~1, and c)~survey data from readers in Part~2.
We transcribed the interview data and analyzed the transcripts through thematic coding and affinity diagrams.
The first author led the qualitative data analysis, and met the research team weekly to discuss findings in order to reduce subjectivity during the data analysis phase
(see Appendix~\ref{appendix:codebook} for the full list of emerging themes). 
\revision{When reporting findings from our qualitative data, we follow~\cite{Abbas_2023} and use the following phrases to indicate the portion of our writer participants who shared each insight: a few (1-5), some (6-10), most (11-15), nearly all (16-19).}

%We used R to perform quantitative data analyses.
For any within-subject comparison (e.g. writers' behavioral logs using the two versions of the AI tool), we used 
%the \verb|lmer| package to run 
linear mixed effect models, controlling for participants' subject IDs and order effect.\footnote{To run the linear mixed effect models, we used the $|lmer|$ R package.}
For scales that capture individuals' opinions toward AI-assisted writing (e.g., readers' survey responses), we used Wilcoxon one-sample tests to examine whether participants' ratings differ significantly from the midpoint (3) of the five-point scales.\footnote{We used Wilcoxon one-sample tests instead of normality tests for distribution as each of these variables was measured by a single 5-point scale. Thus, we treated the data as ordinal instead of continuous numeric values.}
\revision{Details for our analytic approaches are listed in Table~\ref{tab:stats_model}.}
We organize our results according to our research questions with data from different streams, when applicable. We also present two additional points that emerged from our interviews with writers:
\begin{itemize}
    \item In Section~\ref{sec:def_auth}, we first address writer-centered definitions of authenticity in writing and the impact of co-writing with AI on authenticity (RQ1).
    \item In Section~\ref{sec:writing_practice}, we elaborate on how writers would like to preserve their authentic voices through practices of writing with AI (RQ2).
    \item In Section~\ref{sec:personalization}, we compare whether and how personalization of AI writing assistance affects writers' and readers' experiences, respectively (RQ3).
    \item As discussions around writer-reader relationships as well as alternative forms of writing support emerged frequently in our interviews with writers, we further elaborate on them in Section~\ref{sec:writer_reader_relationship} and Section~\ref{sec:future_support}, respectively.
\end{itemize}

\begin{table}[h!]
    \centering
    \renewcommand{\arraystretch}{1.1} 
    \resizebox{0.99\textwidth}{!}{%
    \begin{tabular}{p{0.4\textwidth}|p{0.41\textwidth}|p{0.38\textwidth}}
    \toprule
        Dependent variable & Test goal & Model tested \\
        \hline
        \rowcolor{lightgray!20!}
        \multicolumn{3}{c}{\textbf{Part 1: Writers' Behavioral Data}} \\
        \hline
        Frequency of requesting AI assistance; \newline Acceptance rate of AI suggestions%; \newline Portion of text (by word count) adopted from AI 
        & To compare behavioral patterns of each writer in the personalized vs. non-personalized condition & lmer(DV $\sim$ condition + session order + (1|writerID), data) \\
        \hline
        \rowcolor{lightgray!20!}
        \multicolumn{3}{c}{\textbf{Part 2: Readers' Self-report Data}} \\
        \hline
        Familiarity with generative AI; \newline Interest in reading AI-assisted writing; \newline Authenticity of AI-assisted writing & To understand readers' existing perception toward AI-assisted writing by examining the distribution of each variable & wilcox.test(DV, $\mu= 3$, conf.int=TRUE) \\
        \hline
        Likeability, enjoyment, and creativity of writing; \newline Likelihood of human writing & To compare each reader's perception after reading the three writing passages (writers' solo work, work co-written with personalized AI, and work co-written with non-personalized AI) & lmer(like $\sim$ condition + reading order + (1|authorID) + (1|readerID), data) \\
        \hline
        Degree of preserving writers' authentic voices; \newline Credits and authorship of work & To examine how each reader compare the work co-written with personalized AI vs. non-personalized AI to a writer's solo work respectively & lmer(like $\sim$ condition + reading order + (1|authorID) + (1|readerID), data) \\
        \hline
        Perception of the writing; \newline Perception of the human writer; \newline Appreciation and evaluation of writing & To understand readers' perception toward AI-assisted writing after reading AI-co-written work by examining the distribution of each variable & wilcox.test(DV, $\mu= 3$, conf.int=TRUE) \\
        \bottomrule
    \end{tabular}}
    \caption{Statistical models used for quantitative data analyses}
    \label{tab:stats_model}
\end{table}

\subsection{Writer-Centered Conceptions of Authenticity in Writing}
\label{sec:def_auth}

Our findings first reveal that writers conceptualize authenticity through \textit{the source of content}, \textit{internal experiences and identities} that ground their work, and the actual \textit{writing outcomes}. 
Although participants' reflections did resonate with some of the definitions of authenticity as established in the existing literature (i.e., category, source, and value), they placed further emphasis on viewing authenticity through their internal experiences in addition to through their explicit expressions in writing.
Moreover, regardless of how participants defined and understood authenticity, many of them indicated that authentic writing \textit{is} the essence of good writing, and saw the likely impact of AI on authenticity in writing.
\revision{We provide summaries and quotes of each participant's conceptions of authenticity in Table~\ref{tab:finding_authenticity} and further elaborate on the three key themes of writer-centered definitions of authenticity from Sections~\S\ref{sec:authenticity_source} to \S\ref{sec:authenticity_content}.}

\begin{table}[b!]
    \centering
    \caption{\revision{Writer participants' definitions of authenticity. ``\textit{Source}'', ``\textit{Content}'', and ``\textit{Value alignment}'' are concepts of authenticity identified in prior literature. Construction and expression of authenticity are additional concepts pointed out by our writer participants.}}
    \label{tab:finding_authenticity}
    \resizebox{\textwidth}{!}{%
    \begin{tabular}{p{0.06\textwidth} | p{0.28\textwidth} | p{0.55\textwidth} | p{0.11\textwidth} | p{0.11\textwidth} | p{0.11\textwidth} | p{0.11\textwidth}}
    \toprule
        \textcolor{black}{\textbf{Writer}} & 
        \textcolor{black}{\textbf{Summary of participant's definition of authenticity}} & 
        \textcolor{black}{\textbf{Quote}} & 
        \textcolor{black}{\textbf{Source\newline authenticity (\S\ref{sec:authenticity_source})}} & 
        \textcolor{black}{\textbf{Authentic self \newline(\S\ref{sec:authenticity_self})}} &
        \textcolor{black}{\textbf{Content\newline authenticity (\S\ref{sec:authenticity_content})}} & 
        \textcolor{black}{\textbf{Value alignment}} \\
        \hline
        \textcolor{black}{P1} & 
        \textcolor{black}{Authenticity concerns whether a piece of work is produced by and sounds like its author} & 
        \textcolor{black}{''\textit{To me, authenticity is ensuring that my own ideas are conveyed through the writing, rather than repeating or summarizing another's. [...] Did I come up with the idea? Is this unique? Has it been done before? [...] I want to feel like it's written by [P1's name]}.''} & 
        \textcolor{black}{\checkmark} & &
        \textcolor{black}{\checkmark} & \\
        \hline
        \textcolor{black}{P2} & 
        \textcolor{black}{Authenticity concerns whether a piece of work is produced by and sounds like its author} & 
        \textcolor{black}{''\textit{I only consider it[authenticity] in terms of 'If you're starting to sound too much like this other writer' [...] I guess it's harder to come up with an original idea just because there's only so many ideas in the world. But an authentic voice is always possible.}''} & 
        \textcolor{black}{\checkmark} & &
        \textcolor{black}{\checkmark} &  \\
        \hline
        \textcolor{black}{P3} & 
        \textcolor{black}{Authenticity concerns the human and their living experiences behind the story they wrote.} & 
        \textcolor{black}{``\textit{Can they get inspiration from AI? Absolutely. But authentic writing is about the story that's told from the heart of an author. If there is no human behind the story [...] their experiences, their emotions, their insights, and their imagination behind the story, it wouldn't be as meaningful. And writing would just be pieces of words.}''} & & \textcolor{black}{\checkmark} & & 
         \\
        \hline
        \textcolor{black}{P4} & 
        \textcolor{black}{Authenticity concerns whether the author's beliefs align with their work and whether they feel comfortable having their names on and representing themselves through their work.} & 
        \textcolor{black}{``\textit{Authenticity refers to original work that accurately represents what the author means to convey. What they’re putting out should have their full interest. [...] So I think authenticity has to refer to whether the author could feel strongly standing behind their own work. Whether they feel comfortable having their names behind the work.}''} & & \textcolor{black}{\checkmark} & 
        & 
        \textcolor{black}{\checkmark} \\
        \hline
        \textcolor{black}{P5} & 
        \textcolor{black}{Authenticity concerns \textit{who} produces the work and puts effort into building connection with their readers.} & 
        \textcolor{black}{``\textit{Authentic writing is raw, it creates a human connection between two people: the author and reader. The author builds their soul into their work and trusts the reader to understand them. [...] Authenticity is defined by the source of who produces the writing.}''} & 
        \textcolor{black}{\checkmark} & \textcolor{black}{\checkmark} & & \\
        \hline
        \textcolor{black}{P6} & 
        \textcolor{black}{Authenticity concerns whether the work uniquely represents the author. It is done through the writer's personal touch that is shaped by their experience and worldview.} & 
        \textcolor{black}{``\textit{Authenticity is originality. It's like a fingerprint. It is what makes content unique. As a writer, it's that unique perspective and personal touch every writer brings to a particular topic. [...] Authenticity represents one's personal approaches to their work, which is very much based on their life experiences and understanding of the world.}''} & & 
        \textcolor{black}{\checkmark} & \textcolor{black}{\checkmark}  &\\
        \hline
        \textcolor{black}{P7} & 
        \textcolor{black}{Authenticity represents the writer's unique voice. This is done through expressing the writer's own experiences and sentiment.} & 
        \textcolor{black}{``\textit{It's a certain individual nuance that's just not the same as anyone else. [...] I make it [authenticity] by adding in my own sensibilities. It's the expression of myself and my own emotions.}''} & & 
        \textcolor{black}{\checkmark} & \textcolor{black}{\checkmark} & \\
        \hline
        \textcolor{black}{P8} & 
        \textcolor{black}{Authenticity concerns the writer's own unique voice and style in writing} & 
        \textcolor{black}{``\textit{I believe authenticity in writing is dependent on the writer's ability to present their ideas in their own writing voice, which is completely unique to the individual. To me, as a writer, your voice is the same as your fingerprint. It is what defines you as a writer.}''} & & 
        & \textcolor{black}{\checkmark} & \\
        \hline
        \textcolor{black}{P9} & 
        \textcolor{black}{Authenticity concerns whether the author can express himself/herself freely through their writing.} & 
        \textcolor{black}{``\textit{Authenticity in writing means being able to express one's true self [...] just that I would be able to be myself in my own writing.}''} & & \textcolor{black}{\checkmark} & & \\
        \hline
        \textcolor{black}{P10} & 
        \textcolor{black}{Authenticity is constructed through the living environments and stories of the writer.} & 
        \textcolor{black}{``\textit{Authenticity in writing means to maintain the original stories and voice of the writer. [...] A large part of authenticity comes from the experience and environment that the creators are situated in.}''} & & \textcolor{black}{\checkmark} & & \\
        \hline
        \textcolor{black}{P11} & 
        \textcolor{black}{Authenticity means building connections with the audience. The writer approached this by speaking genuinely to their readers.} & 
        \textcolor{black}{``\textit{To me, authenticity means incorporating my personal voice. I want them to feel like I am talking \emph{to} them [readers], not \emph{at} them. [...] I think just being real and talking to the audience like a person. More just trying to form those relationships and have that more personal connection, put a little personality on when it's appropriate.}''} & 
        \textcolor{black}{\checkmark} & \textcolor{black}{\checkmark} & & \\
        \hline
        \textcolor{black}{P12} & 
        \textcolor{black}{Authenticity concerns writing about subjects that represent one's interest and soul, which makes their work unique from others.} & 
        \textcolor{black}{``\textit{I have ghostwritten articles about many topics. I was aware of my inauthenticity [in] writing about subjects for which I lacked feeling. After several articles in these lanes, I could not find the vocabulary to make a difference from my fellow writers. [...] I want sounds and words and sentences to convey my passion, my thoughts, my soul.}''} & & 
        \textcolor{black}{\checkmark} & \textcolor{black}{\checkmark} & \\
    \bottomrule
    \end{tabular}}
\end{table}
\begin{table}[h!]
    \centering
    \resizebox{\textwidth}{!}{%
    \begin{tabular}{p{0.11\textwidth} | p{0.3\textwidth} | p{0.5\textwidth} | p{0.11\textwidth} | p{0.11\textwidth} | p{0.11\textwidth} | p{0.11\textwidth}}
    \toprule
        \textcolor{black}{\textbf{Writer\newline participant}} & 
        \textcolor{black}{\textbf{Summary of participant's definition of authenticity}} & 
        \textcolor{black}{\textbf{Quote}} & 
        \textcolor{black}{\textbf{Source\newline authenticity (\S\ref{sec:authenticity_source})}} & 
        \textcolor{black}{\textbf{Authentic self \newline(\S\ref{sec:authenticity_self})}} & 
        \textcolor{black}{\textbf{Content\newline authenticity (\S\ref{sec:authenticity_content})}} & 
        \textcolor{black}{\textbf{Value alignment}} \\
        \hline
        \textcolor{black}{P13} & 
        \textcolor{black}{Authenticity arises from the individuality revealed through one's work. It comes from the author expressing himself/herself to produce their work.} & 
        \textcolor{black}{``\textit{I really value the expression of other people. [...] Even if the person isn't good at expressing themselves, you're learning something about another person through that work. You're learning something about what took up space in their mind and in their heart. So there's an authenticity to any kind of human generated art versus computer generated art, that to me, that's the main difference.}''} & 
        \textcolor{black}{\checkmark} & \textcolor{black}{\checkmark} & & \\
        \hline
        \textcolor{black}{P14} & 
        \textcolor{black}{Authenticity is the lived experiences that the author had and could instill in their work.} & 
        \textcolor{black}{``\textit{It [authenticity] is connection to a person's lived experience, a story seen from their point of view. [...] I'm Asian American, so if there was a story written by ChatGPT, and it was about the Asian American experience. The AI could essentially take from sources online to create that, but would it have the true authenticity behind it? And a lot of times authenticity is like your life experience. I could instill the life experiences that I've had to make them [characters in P14's work] feel authentic in their stories.}''} & & \textcolor{black}{\checkmark} & & \\
        \hline
        \textcolor{black}{P15} & 
        \textcolor{black}{Authenticity is expressing oneself through their own unique voice.} & 
        \textcolor{black}{``\textit{Writing authentically involves expressing yourself authentically, from the inside out. And when someone uses their unique, authentic voice, they write best.}''} & & 
        \textcolor{black}{\checkmark} & \textcolor{black}{\checkmark}  & \\
        \hline
        \textcolor{black}{P16} & 
        \textcolor{black}{Authenticity means incorporating and representing one's true self through their work.} & 
        \textcolor{black}{``\textit{Authenticity for me is representing myself through my writing. When people read my work, they should be able to recognize it without viewing my name. [...] It means bringing forth yourself into your writing. Not to mimic others. Just bring you, your personality and everything into your writing.}''} & & 
        \textcolor{black}{\checkmark} & \textcolor{black}{\checkmark} & \\
        \hline
        \textcolor{black}{P17} & 
        \textcolor{black}{Authenticity reflects one's identity through their work. During the process of writing, one might also reshape their own identity.} & 
        \textcolor{black}{``\textit{Authenticity is how your sense of self reflects in your writing. Authenticity doesn't mean being opinionated, rather it’s recognizing your bias, challenging it, and constantly reshaping your worldview through and in your writing.}''} & 
        \textcolor{black}{\checkmark} & \textcolor{black}{\checkmark} & & \\
        \hline
        \textcolor{black}{P18} & 
        \textcolor{black}{Authenticity is the true self and soul that stand behind a piece of writing.} & 
        \textcolor{black}{``\textit{Authenticity just comes down to having that soul behind the writing to make it feel like it's compelling and being written by someone who's passionate about what they're writing about. [...] Just having that level of compassion and understanding from a person's perspective, I think makes a big difference in writing.}''} & & \textcolor{black}{\checkmark} & & \\
        \hline
        \textcolor{black}{P19} & 
        \textcolor{black}{Authenticiy arises from the individual experiences that allow one to tell stories in their own way.} & 
        \textcolor{black}{``\textit{Authenticity is your signature, your personalized way of telling a story. [...] Everyone can start with similar ideas but individual experiences allow for personal, unique ways of storytelling.}''} & & \textcolor{black}{\checkmark} & & \\
    \bottomrule
    \end{tabular}}
\end{table}

\subsubsection{Defining authenticity through the source of content (Source authenticity)}
\label{sec:authenticity_source}
% This point is related to the construct of “source” as in prior literature
% Impact of co-writing with AI on authenticity: When AI participates in writing, writers are no longer the sole source of content generation, leading questions to authenticity in writing.

Several writers described authenticity as a \textit{``who''} question, focusing on who wrote the text or was the source of the content.
This concept mirrors \textit{Source}, a long-standing theoretical component of authenticity in the existing literature, and is directly related to both writers' and readers' considerations for authorship.
\revision{Participants whose conceptualizations aligned with \textit{Source} authenticity emphasized the \textit{entity} who took actions and contributed to a piece of work. For example, such actions might include producing a piece of text or trying to understand the audience's interest.}

Writers who held this view also saw AI writing assistance as a direct threat to authenticity.
With AI participating in the writing process, writers are no longer the sole source of content generation, raising questions about the authenticity of their writing.
\revision{A few participants highlighted the importance of having writers produce their work themselves, as they learn, revise, and refine their work through the actions and processes of writing. Furthermore, they also believed that audiences learn more about writers through the way they produce content.}

\begin{comment}
\begin{quote}
    \revision{``\textit{I'd almost rather see a random person on the street draw a tree or write a story about their day than get that writing from an artificial intelligence. Even if the person isn't good at expressing themselves, you're learning something about another person through that work. You're learning something about the way that they view a tree based on the way that they draw it. [...] %Or you're learning something about what they ... if you ask them to write a story about their day and the whole page is focused on one interaction that they had, 
    Even if it's poorly written, you're learning something about what took up space in their mind and in their heart. So there's an authenticity to it that to any kind of human-generated art versus computer-generated art, that's the main difference.}'' (P13)}
\end{quote}
\end{comment}

%\subsubsection{Defining authenticity through the process and writers' internal experiences of constructing and expressing their authentic self}
\subsubsection{Defining authenticity through constructing and expressing their authentic self (Authentic self)}
\label{sec:authenticity_self}

Writers mentioned several internal states during their writing processes as key constructs of authenticity, many of which have been less covered by prior literature. 
%not yet been thoroughly discussed in the existing literature.
These include (1) whether writers can freely express their emotions to form emotional connections with their readers, (2) whether the process of writing allows them to feel passionate about their work, (3) whether they have the autonomy to select topics and content that are personally important to them, (4) and most importantly, whether a writer can justify having their name and identity behind their work.
In other words, a writer claims the authenticity of their writing if they can soundly argue why the piece of work can only be done by them as \textit{the} writer.

This view of authenticity is shaped by the belief that writers’ identities, backgrounds, and lived experiences serve as fundamental materials of writing and enrich their work.
These cornerstones---whether writers can genuinely express their own experiences and identities through the writing as well as whether they feel enthusiastic and believe in the importance of
%would feel importance and enthusiasm for 
their work---jointly contribute to authenticity.
% ---due to these personal values that are attached to their work
As such, the work would never be in its current form if produced by any other writer.
One writer described it metaphorically:

\begin{quote}
    ``\textit{I've often said that writing is a lot like a tree. The trunk of the tree is the idea. Everybody can have the same idea and then start to write something. But eventually, once you get up into the branches, everybody's going to go off in their own different branch. Authenticity is your own personal branch, the way that you would take an idea and how it would be different from 50 other people who take the same idea and try to finish it or complete it to the end.}'' (P19)
\end{quote}

However, when AI contributes to writing, the content is no longer grounded solely in writers’ own experiences, memories, backgrounds, knowledge, research, and more.
It is not always clear to them where the writing suggestions were coming from and what they were grounded on, and thus, writers can no longer claim their work as a reflection of their lived experiences or as representative of themselves.

\subsubsection{Defining authenticity through writing outcomes (Content authenticity)}
\label{sec:authenticity_content}
% This point is related to the construct of “category” as in prior literature, but it is also the least prominent for writers
% Impact of co-writing with AI on authenticity: AI’s influences might shift writing output from the “typical” ways in how writers write.

Examining the outcomes of writing is yet another approach to assessing authenticity.
More specifically, authentic writing equals writing that best represents the work of a writer, where one can tell who the writer is by reading the text.
This perspective echoes the ``category'' construct of authenticity, which is the most adopted definition of authenticity in the literary studies space~\cite{Auth_Humanity_Guignon_2008, Auth_Humanity_AMA_Lehman_2019, Auth_Humanity_Handler_1986}.
Nonetheless, it was the least referred to and a less critical construct from writers' points of view.
In our study, the few writers who defined authenticity through this lens were concerned about the influences of AI writing assistance on authenticity, as AI's input might ``shift writers away from the typical ways of [their] writing practices'' (P2), intruding on their usual tones, voices, and ways of presentation in their writing.
Under this conception of authenticity, since authenticity in writing is evaluated by comparing it to a writer's representative work, the writers in our study also connected the meaning of authenticity back to various writing elements (e.g., word choice, style, use of references and metaphors) that uniquely characterize their own writing.

\subsection{(Re)claiming Authenticity Through Practices of Co-writing with AI}
\label{sec:writing_practice}

Through the co-writing sessions, writers identified several practices during the writing process that could help shape authenticity in writing with AI assistance.
In other words, \textit{how} writers use AI while writing plays a key role in determining the authenticity of AI-assisted writing. 
These determinants include: (1) whether one starts with a clear vision for writing in mind, (2) when one calls for writing assistance from AI, (3) what portion of contribution one makes in the writing, and (4) what purpose and usage that AI writing assistance serves.

%\subsubsection{Is there an image at the starting point?}
\subsubsection{Setting off with a clear vision of what to achieve in one's writing}
% whether writers had a clear idea or a clear image of what they want and how they want to present themselves in writing

In describing their writing, nearly all writers reported that they seldom start writing with a blank sheet.
Instead, they typically begin their processes with some vision for writing in mind.
Such a vision need not be a concrete, fully fleshed-out idea. 
But writers often already have certain directions or settings in mind that they would like to set up for the passage, given the preparatory work they did before writing (e.g., ideation, research, collecting references and other writing materials). 
Otherwise, they lack an anchor to cohesively guide how they respond to, select, and curate AI suggestions alongside their own writing.
In such a case, writers worried that they might be at more risk of being affected by AI suggestions, depriving their authentic voices in writing.

\begin{quote}
    ``\textit{I think it [AI] has the possibility to [affect writers] if you let it. But if you already have your ideas of what you want to stay true to, then it isn't going to affect your authentic voice, in my opinion. So to me, it comes down to user influence and how much the user chooses the directions that the AI suggests.}'' (P4)
\end{quote}

%\subsubsection{When did writers use AI during their writing processes?}
\subsubsection{Working with AI through the ``fuzzy area'' of idea development when writing}
% there is a fuzzy area in the writing process, where one already has a broad idea/direction for writing but it hasn't been externalized, shaped, and refined into an actual piece of writing
% this is where AI has the most potential to support writing without compromising authenticity

It is worth recognizing that writers work through various stages throughout their writing processes.
While writers bring a premature vision to initiate writing, there is typically a ``fuzzy area'' (P04) between the starting point and well-developed ideas, which eventually leads to a fully crafted piece of writing.
Most writers believe that AI writing assistance is most acceptable and is considered as having the least harm on authenticity at this middle ground for two reasons.
First, the ways in which writers would interact with AI writing assistance tools during this stage primarily serve to consolidate and further develop their ideas (we discuss opportunities for AI tools to support writers more extensively in Section~\ref{sec:AI_usage}).
At this stage, writers write to incubate and organize their thinking rather than to produce text to construct the actual piece of writing.
Therefore, though AI suggestions contribute pieces of content, they serve as materials that facilitate the thought processes rather than the writing \textit{per se}.

% \subsubsection{Who did more work for a piece of writing?}
\subsubsection{Claiming contribution through content gatekeeping}
% Evaluating portions of contributions
% Human writers as content gatekeepers

Most importantly, in our study, most writers considered AI-assisted writing as authentic as long as they contributed more to the writing.
Specifically, writers noted that one's \textit{contribution} does not necessarily equal how much text they wrote. %distinguished the portion of \text{contribution} from the actual length of text written. 
In their view, real contributions require ``content gatekeeping'' (P8)---that is, actively deciding what goes into the writing content.
Given this conception of contribution, writers frequently described themselves as doing most of the work during the writing process as they took charge of selecting, revising, and incorporating AI suggestions into their writing.
Whether a word, a sentence, or a paragraph is produced with AI assistance or not, it is the human writer, instead of the AI tool, that takes control over whether to adopt, remove, or revise each piece of content.

% \subsubsection{What did writers use AI for?}
\subsubsection{Beyond AI as a co-writer: Opportunities for supporting authentic writing processes} %\VL{I am struggling a bit with what this section is about...the content does not always suggest concrete "co-writing forms'', and they are not "beyond writing". I first tried to revise the title as ``beyond content suggestions'', but I saw at least the first two points are still about content suggestion somewhat...I feel they are "(high-level) conceptualization of how AI can contribute to (authentic) writing process'' or ``AI opportunities for authentic writing processes''...} \angel{so kind of close to the idea of "beyond writing"... this group here says about how AI can help to preserve writers' authentic voices in writing but *not* by doing the writing job, especially not doing the job of producing a full piece of text.} \VL{When you say ``beyond writing'', it is not clear this is about ``AI doing the writing'', what about ``beyond AI as a co-writer/coauthor: opportunities for supporting authentic writing processes''?}
\label{sec:AI_usage}
% AI as internalization tool
% AI as jumping points and driving forces that keep the writing flow going
% AI as sounding board

Besides requesting content suggestions from AI (i.e., what the CoAuthor interface supports), some writers believed they could benefit from AI assistance while preserving their authentic voices by leveraging AI capabilities as  \textit{a means of internalization}, \textit{a driver of the writing flow}, and \textit{a sounding board of public feedback}. Below, we unpack each of these favorable co-writing forms that our participants indicated:

\textbf{\textit{AI as a means of internalization:}}
Nurturing themselves through reading, researching, and experiencing the world is key to the quality and richness of writers' work. 
Writers often improve their work over time by reading and getting inspired by others' work.
Thus, some writers saw AI as a means to help them absorb large amounts of information and saw AI suggestions as excerpts resulting from this practice.
As P18 described: 

\begin{quote}
    ``\textit{The way that we actually improve ourselves is by reading other people's works. Our memory retention on how other people worded things is in the background of our minds when we need to write something of a similar nature. [...] But really what you're doing is regurgitating what you've read maybe 20 years ago, ten years ago, or just five or 4 hours ago, doing it in your own unique way and telling your own unique story. % Stephen King even promotes the idea of reading for 4 hours a day and writing for 4 hours a day. 
    [...] This [AI writing assistance] is kind of combining those two things [reading and writing] all at once. So instead of reading, I'm getting an option which is exactly what the reading is supposed to do for you. And I can choose that option or choose to go with my own authenticity, my own way of actually writing it.}''
\end{quote}

Under this form of AI assistance, some participants saw an analogy with how AI is built (i.e., learning from large amounts of text to produce text): much as human writers consume information and experiences to enrich their writing, AI digests vast amounts of data to generate text---though the machine expedites this process significantly.
In other words, while writers gradually take in their work and life experiences to enrich their writing, they saw AI as a potential tool that compresses this process and presents the results of some sort of ``internalization.''
As P10 put it in a critical way, ``if what this AI does for the writer is unauthentic, then, perhaps there is no such thing as authentic writing.''

\textbf{\textit{AI as a driver of the writing flow:}}
Having the option to continuously request assistance from AI may help carry on writers' writing flow and remove writing blocks.
Occasionally, AI suggestions might point to novel directions or ideas that writers would not have conceived of by themselves, and the tool might also offer ``jumping points'' that allow writers to transition from one idea to another. 
However, writers valued the AI's apparent capability of ``keeping [writers] up with their momentum'' (P08) the most.
Not only does it take warm-up time for writers to immerse themselves in a smooth writing flow, but once the flow experience is interrupted, it can be difficult to resume.
Writers saw AI suggestions as on-demand, temporary remedies when they sensed such an interruption in their writing process.

\textbf{\textit{AI as a sounding board of public feedback:}}
Several writers were aware that large language models are trained with vast amounts of text data across digital outlets.
As such, they viewed these models as an assembly of information and opinions online, and would like to use it as a sounding board to potentially project and reflect responses from broader audiences.
Some writers saw AI suggestions as syntheses of the public's interests, and they were interested in using such information to navigate and/or improve their own work.

\subsection{Finding and Preserving Authentic Voice: Personalization as Double-Edged Sword}
\label{sec:personalization}
\revision{While most writers showed preferences toward personalized AI writing assistance, they foresaw both positive and negative influences of personalization. They believed grappling with such dilemmas was key to adopting AI tools in what they dubbed as a more collaborative rather than reliant fashion.}
In this section, we discuss how writers perceived the impact of personalized AI on their writing, together with whether and how personalization affected their usage behaviors with the tool, and how readers responded to their writings with personalized versus non-personalized AI assistance based on data from the reader survey.

\begin{table*}[b]
    \centering
    \renewcommand{\arraystretch}{1.2} 
    \resizebox{\linewidth}{!}{
    \begin{tabular}{p{0.35\textwidth}|c c c c |c c c c |c c c c}
        \toprule
        \textbf{Variable} & \multicolumn{4}{c|}{\textbf{Round 1}} & \multicolumn{4}{c|}{\textbf{Round 1 (removed 7 responses)}} & \multicolumn{4}{c}{\textbf{Round 2}} \\
        \hline
        \rowcolor{lightgray!20!}
        \multicolumn{13}{c}{\textbf{Opinions about AI writing assistance}} \\
        \hline
        & \multicolumn{2}{c}{$Z$} & \multicolumn{2}{c|}{$p$} & 
        \multicolumn{2}{c}{$Z$} & \multicolumn{2}{c|}{$p$} & 
        \multicolumn{2}{c}{$Z$} & \multicolumn{2}{c}{$p$} \\
        \hline
        \textbf{Familiarity with generative AI} & 
        \multicolumn{2}{c}{$5.21$} & \multicolumn{2}{c|}{$<0.001^{***}$} & 
        \multicolumn{2}{c}{$5.63$} & \multicolumn{2}{c|}{$<0.001^{***}$} & 
        \multicolumn{2}{c}{$5.54$} & \multicolumn{2}{c}{$<0.001^{***}$} \\
        \hline
        \textbf{Interest in reading AI-assisted writing} & 
        \multicolumn{2}{c}{$4.13$} & \multicolumn{2}{c|}{$<0.001^{***}$} & 
        \multicolumn{2}{c}{$5.47$} & \multicolumn{2}{c|}{$<0.001^{***}$} & 
        \multicolumn{2}{c}{$5.82$} & \multicolumn{2}{c}{$<0.001^{***}$} \\
        \hline
        \textbf{Authenticity of AI-assisted writing} & 
        \multicolumn{2}{c}{$5.38$} & \multicolumn{2}{c|}{$<0.001^{***}$} & 
        \multicolumn{2}{c}{$2.58$} & \multicolumn{2}{c|}{$0.010^{*}$} & 
        \multicolumn{2}{c}{$4.21$} & \multicolumn{2}{c}{$<0.001^{***}$} \\
        \hline
        \rowcolor{lightgray!20!}
        \multicolumn{13}{c}{\textbf{Reading writing samples}} \\
        \hline
        & $\beta$ & $S.E.$ & $t$ & $p$ & 
        $\beta$ & $S.E.$ & $t$ & $p$ & 
        $\beta$ & $S.E.$ & $t$ & $p$ \\
        \hline
        \textbf{Likeability} & & & & & & & & & & & & \\
        Solo vs. Personalized AI & 
        $-0.25$ & $0.20$ & $-1.26$ & $0.215$ & 
        $-0.15$ & $0.16$ & $-0.9$0 & $0.371$ & 
        $-0.24$ & $0.15$ & $-1.56$ & $0.125$ \\
        Solo vs. Non-personalized AI & 
        $-0.05$ & $0.20$ & $-0.25$ & $0.803$ & 
        $-0.09$ & $0.16$ & $-0.52$ & $0.990$ & 
        $-0.15$ & $0.15$ & $-0.97$ & $0.335$ \\
        \hline
        \textbf{Enjoyment} & & & & & & & & & & & & \\
        Solo vs. Personalized AI & 
        $-0.15$ & $0.21$ & $-0.71$ & $0.483$ & 
        $-0.11$ & $0.18$ & $-0.61$ & $0.543$ & 
        $-0.18$ & $0.16$ & $-1.10$ & $0.275$ \\
        Solo vs. Non-personalized AI & 
        $-0.15$ & $0.21$ & $-0.70$ & $0.483$ & 
        $-0.04$ & $0.18$ & $-0.20$ & $0.839$ & 
        $-0.12$ & $0.16$ & $-0.74$ & $0.465$ \\
        
        \hline
        \textbf{Creativity} & & & & & & & & & & & & \\
        Solo vs. Personalized AI &
        $0.91$ & $0.25$ & $0.01$ & $0.999$ & 
        $-0.11$ & $0.22$ & $-0.50$ & $0.619$ & 
        $-0.18$ & $0.19$ & $-0.92$ & $0.360$ \\
        Solo vs. Non-personalized AI & 
        $-0.45$ & $0.24$ & $-1.82$ & $0.077$ & 
        $-0.22$ & $0.22$ & $-1.00$ & $0.322$ & 
        $-0.29$ & $0.19$ & $-1.54$ & $0.129$ \\
        \hline
        \rowcolor{lightgray!20!}
        \multicolumn{13}{c}{\textbf{Comparing all writing samples}} \\
        \hline
        & $\beta$ & $S.E.$ & $t$ & $p$ & 
        $\beta$ & $S.E.$ & $t$ & $p$ & 
        $\beta$ & $S.E.$ & $t$ & $p$ \\
        \hline
        \textbf{Likelihood of human writing} & & & & & & & & & & & & \\
        Solo vs. Personalized AI & 
        $-0.79$ & $0.35$ & $-2.25$ & $0.028^{*}$ & 
        $-0.77$ & $0.28$ & $-2.72$ & $0.008^{**}$  & 
        $-0.73$ & $0.25$ & $-2.94$ & $0.004^{**}$ \\
        Solo vs. Non-personalized AI & 
        $-0.32$ & $0.35$ & $-0.90$ & $0.372$ & 
        $-0.54$ & $0.28$ & $-1.90$ & $0.061$ & 
        $-0.64$ & $0.25$ & $-2.57$ & $0.012^{*}$ \\
        \hline
        \rowcolor{lightgray!20!}
        \multicolumn{13}{c}{\textbf{Comparing personalized vs. non-personalized writing samples}} \\
        \hline
        & $\beta$ & $S.E.$ & $t$ & $p$ & 
        $\beta$ & $S.E.$ & $t$ & $p$ & 
        $\beta$ & $S.E.$ & $t$ & $p$ \\
        \hline
        \textbf{Preserving writers' authentic voices} & & & & & & & & & & & & \\
        Personalized AI vs. Non-personalized AI &
        $0.40$ & $0.27$ & $1.47$ & $0.164$ & 
        $0.05$ & $0.24$ & $0.19$ & $0.853$ & 
        $0.07$ & $0.21$ & $0.36$ & $0.722$ \\
        \hline
        \textbf{Credits and authorship of work} & & & & & & & & & & & & \\
        Personalized AI vs. Non-personalized AI & 
        $0.55$ & $0.25$ & $2.24$ & $0.037^{*}$ & 
        $0.22$ & $0.23$ & $0.97$ & $0.340$ & 
        $0.18$ & $0.19$ & $0.97$ & $0.339$ \\
        \hline
        \rowcolor{lightgray!20!}
        \multicolumn{13}{c}{\textbf{Opnions after reading AI-assisted writing}} \\
        \hline
        & \multicolumn{2}{c}{$Z$} & \multicolumn{2}{c|}{$p$} & 
        \multicolumn{2}{c}{$Z$} & \multicolumn{2}{c|}{$p$} & 
        \multicolumn{2}{c}{$Z$} & \multicolumn{2}{c}{$p$} \\
        \hline
        \textbf{Perception of the writing} & 
        \multicolumn{2}{c}{$2.68$} & \multicolumn{2}{c|}{$0.007^{**}$} & 
        \multicolumn{2}{c}{$2.68$} & \multicolumn{2}{c|}{$0.007^{**}$} & 
        \multicolumn{2}{c}{$3.70$} & \multicolumn{2}{c}{$<0.001^{***}$} \\
        \hline
        \textbf{Perception of the human writer} & 
        \multicolumn{2}{c}{$2.75$} & \multicolumn{2}{c|}{$0.006^{**}$} & 
        \multicolumn{2}{c}{$3.41$} & \multicolumn{2}{c|}{$<0.001^{***}$} & 
        \multicolumn{2}{c}{$4.58$} & \multicolumn{2}{c}{$<0.001^{***}$} \\
        \hline
        \textbf{Appreciation \& evaluation of writing} &
        \multicolumn{2}{c}{$-0.39$} & \multicolumn{2}{c|}{$0.703$} & 
        \multicolumn{2}{c}{$-0.25$} & \multicolumn{2}{c|}{$0.807$} & 
        \multicolumn{2}{c}{$-0.30$} & \multicolumn{2}{c}{$0.763$} \\
        \bottomrule
    \end{tabular}}
    \caption{Results of statistical tests for numeric variables in Part~2 ($p^{*}<0.05$, $p^{**}<0.01$, $p^{***}<0.001$)}
    \label{tab:part2_analysis}
\end{table*}

\subsubsection{Writers' subjective preferences toward personalized AI}
Comparing their experiences co-writing with the personalized AI and the non-personalized AI, writers reflected on both the positive and negative impact of personalization on authentic writing after we revealed that one of the tools they experienced was powered by an LLM personalized with the writing sample they provided.
Overall, in our study, the majority of writers preferred working with personalized writing tools when they were asked to compare the two options.
This is because writers believed personalization could help preserve their genuine voice, express themselves naturally, and better connect with their own identities.

\revision{In this regard, participants identified two main positive outcomes.
First, many participants noted they produced better quality of work under time constraints when working with personalized AI; it reduced the need for going back and forth to adjust and align the generated text with their own voices and allowed participants to focus on producing new content.
Generated output that was viewed as higher quality also yielded more inspiration for writers. As P15 suggested a stark contrast in their comment:}

\begin{quote}
    \revision{``\textit{With the first one [working with personalized AI writing assistance] ... it almost seems like when you have a partner that energizes you and you get inspiration from each other. %Well, I know I can't inspire AI, but yeah, I felt more like that with the first round. 
    In the second round [working with non-personalized AI writing assistance], I felt like [...] almost like you have someone that gives a lot of suggestions that you don't really like, but it didn't really necessarily light me up.}''}
\end{quote}

\revision{In the same vein, some participants expressed a stronger sense of what they described as akin to \textit{collaboration} when working with personalized AI. 
When working with a non-personalized AI, participants more often saw the need to switch to a supervising role to oversee and ensure the passage presented a consistent style. Such experiences were described as ``a solo thing'' (P05). By contrast, participants noted that it felt more like some type of ``bilateral exchange'' when they could focus on throwing out raw content---like the suggestions generated by the AI tool.}

\revision{On the negative end, participants worried personalization might also lead to writers adopting more suggestions from AI, allowing more influences from the tool.
Based on their subjective reflections, many participants believed that they contributed much more to the written content when working with a non-personalized AI, as they more frequently experienced the need to revise content generated by non-personalized AI, leaving more limited room for AI to influence them.
In particular, more experienced writers were concerned that novice writers were more likely affected by frequently adopting suggestions from personalized AI, as they might not yet have established their own styles and voices in written work.}

\subsubsection{Writers adopted a similar degree of assistance from both personalized and non-personalized AI tools.}
Despite writers' subjective preferences for the personalized tool, in our study there was no significant difference in the frequency of requesting AI assistance when working with the personalized vs. the non-personalized tool ($\beta = 0.13$, $S.E. = 1.08$, $t = 0.12$, $p = 0.904$).
Likewise, writers' behavioral data from the writing logs showed no significant difference in the rate of accepting suggestions from AI between the personalized vs. non-personalized condition ($\beta = -0.02$, $S.E. = 0.07$, $t = -0.27$, $p = 0.790$). (Also see Appendix~\ref{appendix:writing_log} for detailed statistics for each writer.)

\begin{comment}
\begin{figure}[h!]
    \centering
    \includegraphics[width=0.9\textwidth]{fig/part1_get.png}
    \caption{Number of times each writer participant pulled up AI assistance when writing with the personalized vs. non-personalized tools.}
    \label{fig:part1_get}
\end{figure}

\begin{figure}[h!]
    \centering
    \includegraphics[width=0.9\textwidth]{fig/part1_accept_rate.png}
    \caption{Rate (\%) of AI suggestions accepted by each writer participant when writing with the personalized vs. non-personalized tools.}
    \label{fig:enter-label}
\end{figure}    
\end{comment}

\subsubsection{Readers' responses to personalized and non-personalized AI-assisted writing}
%From readers' perspectives, they 
The readers in the second part of our study reported similar experiences reading the three types of passage---writers' solo work, the work they produced with personalized AI, and the work with non-personalized AI.
Per readers' numeric ratings, the degree of enjoyment ($F = 0.63$, $p = 0.536$), likeability ($F = 1.23$, $p = 0.298$), and creativity ($F = 1.19$, $p = 0.311$) after reading each type of passage showed no significant difference across the three conditions. Pair-wise comparison also showed no significant difference \revision{(See Table~\ref{tab:part2_analysis} for statistics)}. Furthermore, when asked to select portions of text they believed to be generated by AI, readers were not able to precisely identify which part of the text contained AI-assisted output. The rate of correct identification has no significant difference between the personalized and non-personalized conditions.

\begin{figure}[h!]
    \centering
    \includegraphics[width=0.54\textwidth]{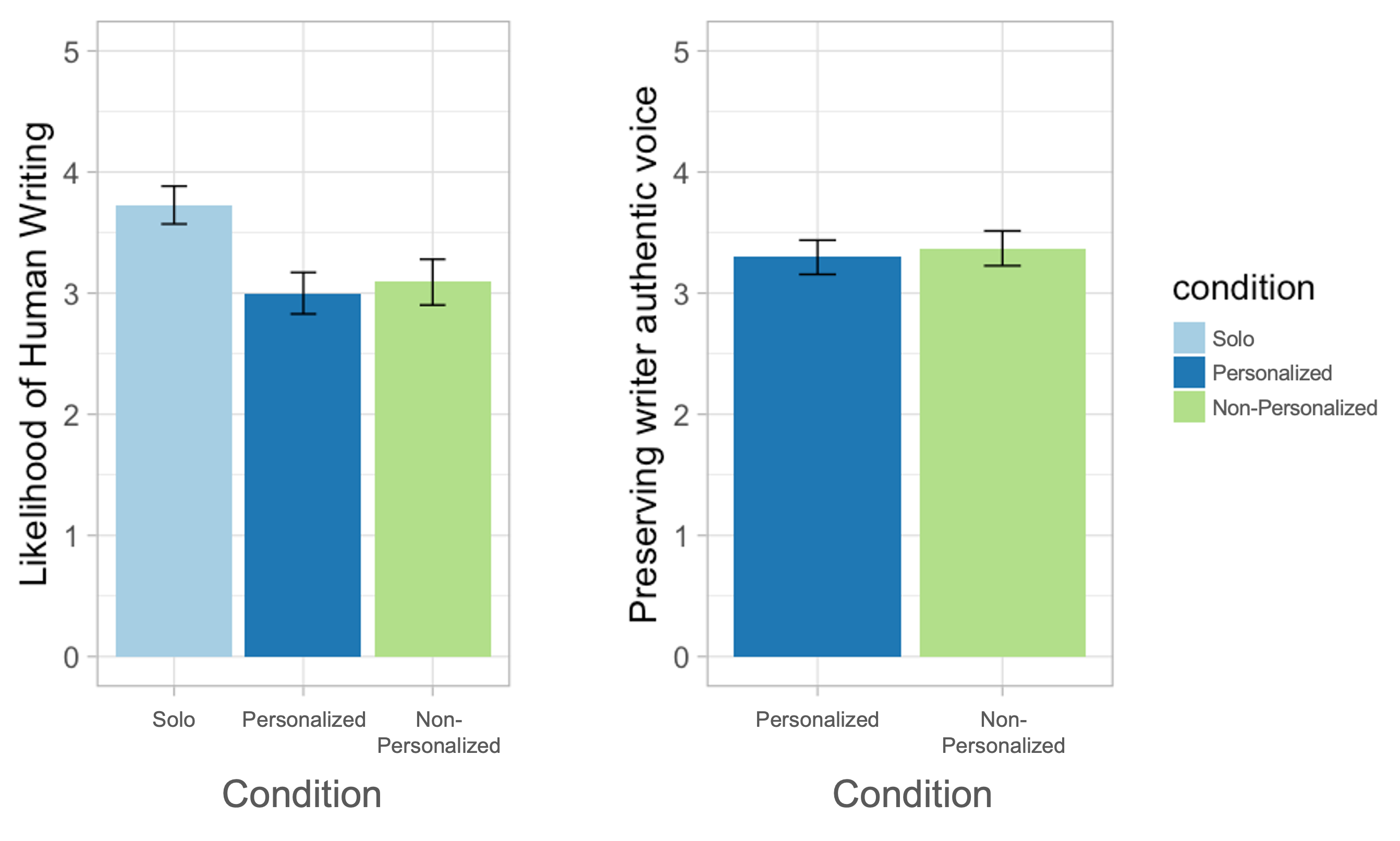}
    \caption{Reader participants' ratings to compare the three pieces of writing. Left: Ratings for whether a piece of writing is likely to be done independently by a human writer (rating = 5) or co-written with AI (rating = 1) on a 5-point Likert scale. Right: Ratings for whether a piece of AI-assisted writing preserves a writer's authentic voice.}
    \label{fig:part2_rating}
\end{figure}

\begin{figure}[h!]
    \centering
    \includegraphics[width=\textwidth]{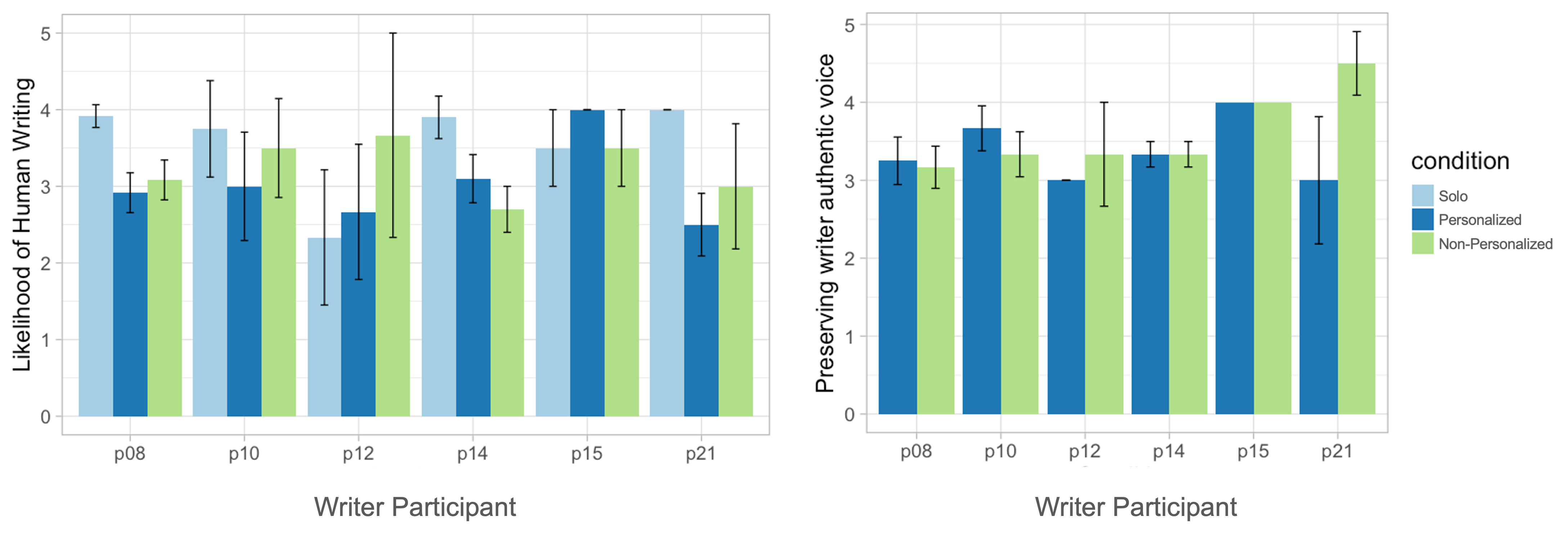}
    \caption{Reader participants' ratings for writing passages for each writer. Left: Ratings for whether a piece of writing is likely to be done independently by a human writer (rating = 5) or co-written with AI (rating = 1) on a 5-point Likert scale. Right: Ratings for whether a piece of AI-assisted writing preserves a writer's authentic voice.}
    \label{fig:part2_prob_human}
\end{figure}

When asked to compare across the three writing passages, our reader participants rated the solo human work as more likely to be done independently by a human writer than the two forms of AI-assisted writing (solo $-$ personalized AI: $\beta = -0.73$, $S.E. = 0.25$, $t = -2.94$, $p = 0.004$; solo $-$ non-personalized AI: $\beta = -0.64$, $S.E. = 0.24$, $t = -2.57$, $p = 0.011$), while there is no significant difference between the work co-written with the personalized vs. non-personalized AI ($\beta = 0.09$, $S.E. = 0.26$, $t = 0.35$, $p = 0.727$).
When comparing the two AI-assisted writing pieces to each writer's solo work, readers did not notice any difference regarding whether the AI-assisted work preserved the writer's authentic voice ($\beta = 0.07$, $S.E. = 0.21$, $t = 0.36$, $p = 0.722$). It is also worth noting that, on average, reader participants rated both versions of AI as ``preserving the writer's authentic voice'' from ``moderate'' (3 on a 5-point scale) to ``a lot'' (4 on a 5-point scale).
\revision{Figures~\ref{fig:part2_rating} and \ref{fig:part2_prob_human} also further illustrate the differences between participants' ratings for the conditions.}

\begin{comment}
\begin{figure}[h!]
    \centering
    \includegraphics[width=\textwidth]{fig/part2_bar_combined.png}
    \caption{Reader participants' ratings for whether a piece of writing is likely to be done independently by a human writer (rating = 5) or co-written with AI (rating = 1) on a 5-point Likert scale.}
    \label{fig:prob_human}
\end{figure}
\end{comment}

We took a closer look at readers' ratings for writing from each of the six writers respectively. For 4 out of the 6 writers, work co-written with non-personalized AI, compared to their work co-written with personalized AI, was perceived as more likely done by a human writer independently. The two writers (P14 and P15) whose work was rated as more likely to be written by a human when they worked with the personalized AI wrote about romance and spiritual topics, respectively. When we cross-checked the ratings from the readers with these writers' behavioral data, we also noticed that these two writers accepted more suggestions from the personalized AI than from the non-personalized during the writing sessions (P14 accepted 67\% of suggestions from the personalized AI vs. 41\% from the non-personalized AI; P15 accepted 71\% suggestions from the personalized AI vs. 20\% from the non-personalized AI).

We note that recent research has found that people could not reliably distinguish whether a piece of content is generated by people or the latest LLMs~\cite{jakesch2023human}. While our empirical results suggest that some readers might be able to tell if a piece is co-written with AI with some degree of confidence, they may not respond differently to content written by the writer alone or content co-written with AI in terms of how much they enjoy reading it, at least not in a ``leisurely reading'' setting like ours where readers encounter a piece of reading from an unfamiliar writer instead of actively seeking a particular writer's work.

%did not have prior familiarity with the writer's work nor actively seek out a particular writer's work.
%at least in the kind of leisurely reading context we studied (i.e. the reader did not have frequent exposure to or actively seek out the writer's work).

\begin{comment}
    \begin{figure}[h!]
    \centering
    \includegraphics[height=0.35\textwidth]{fig/part2_prob_human_author.png}
    \caption{Reader participants' ratings for whether a piece of writing is likely to be done independently by a human writer (rating = 5) or co-written with AI (rating = 1) on a 5-point Likert scale.}
    \label{fig:part2_prob_human}
    \end{figure}
\end{comment}

\begin{comment}
    \begin{figure}[h!]
    \centering
    \includegraphics[height=0.35\textwidth]{fig/part2_cowriting_auth_author.png}
    \caption{Reader participants' ratings for whether a piece of AI-assisted writing preserves a writer's authentic voice.}
    \label{fig:part2_prob_human}
    \end{figure}
\end{comment}

%\subsection{About Time: Rethinking Writer-Reader Relationships in the Age of Generative AI}
\subsection{Rethinking Writer-Reader Relationships in the Age of Generative AI}

\label{sec:writer_reader_relationship}

\subsubsection{Writers were concerned about the devaluation of their work as a result of co-writing with AI.}
Throughout the study, writers expressed concerns about audiences' reactions to their use of AI assistance for their writing.
While some believed readers might have diverging opinions depending on their level of acceptance of new technology, particularly AI, all of the writers in our study expected to receive some negative feedback from readers.
Notably, in writers' view, ``readers'' include not only general audiences who might read and purchase their work but also the clients who directly commission their writing jobs.
They were worried that their use of AI writing assistance---and the fact that anyone can use AI for text production---would result in lower perceived value and misperceptions of their writing as ``easy work'' (P13). 
Nearly all writers expressed increased concerns as adopting AI writing assistance has become more commonplace across various industries.
In the long run, they believed professional writers would continue to appreciate the work of their peers, but they were uncertain whether non-experts, including some of their clients, would change their views about the value of human writing as AI-writing assistance becomes widely accessible. 
% \placeholder{mention something about the writer economy + authenticity as value for pay}

\subsubsection{Readers appreciated writers' attempt to explore new technology and expressed interest in reading AI-co-created text.}
\begin{figure}[h!]
    \centering
    \includegraphics[height=0.35\textwidth]{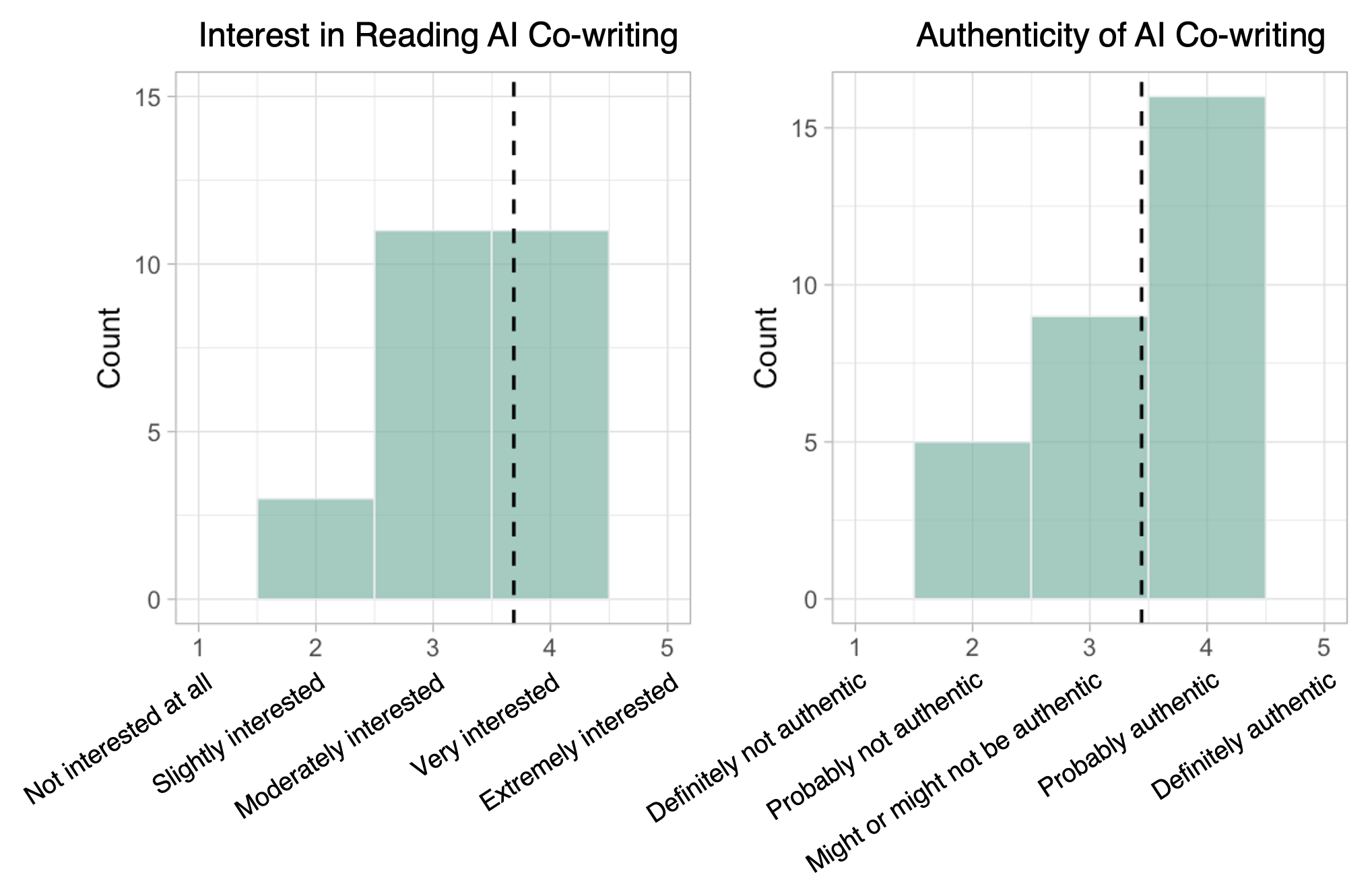}
    \caption{Readers' perceptions toward human-AI co-writing \textit{before} reading AI-assisted writing. Dashed lines represent mean values.}
    \label{fig:part2_pre_perception}
\end{figure}

By contrast, readers in our study held a more positive view toward the use of AI writing assistance. Before actually reading these AI-co-written passages, readers' levels of interest are already significantly skewed toward the positive end ($Z = 5.82$, $p < 0.001$) \revision{(See Figure~\ref{fig:part2_pre_perception} for the distribution)}.
The majority of them also believed work co-written with AI should be viewed as authentic writing as well ($Z = 4.21$, $p < 0.001$). Readers also suggested that whether a piece of work was co-written with AI would not affect how they evaluated and appreciated the writing ($Z = -0.30$, $p = 0.763$).

\begin{figure}[h!]
    \centering
    \includegraphics[height=0.36\textwidth]{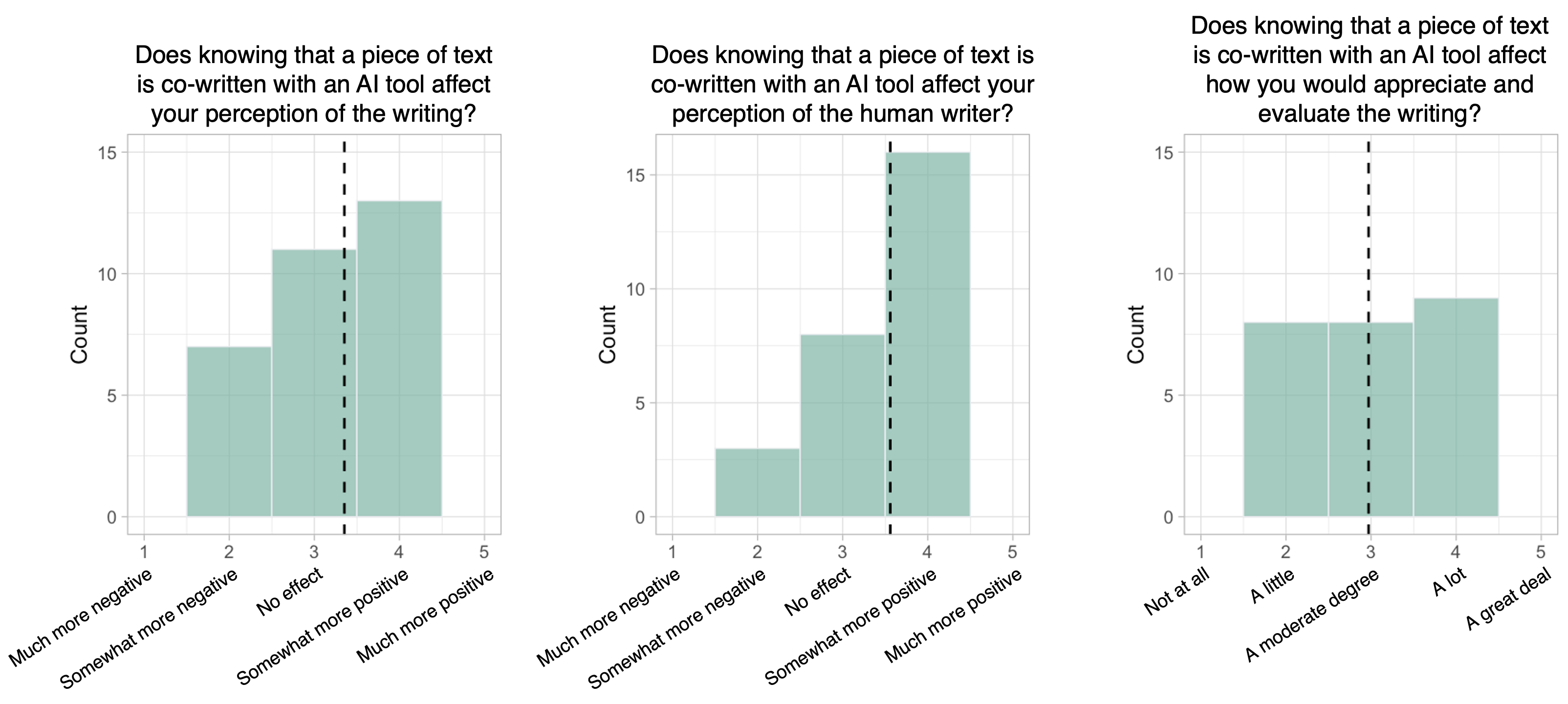}
    \caption{Readers' perceptions toward the writing, the writer, and their approaches to evaluating work \textit{after} reading AI-assisted writing. Dashed lines represent mean values.}
    \label{fig:part2_post_perception}
\end{figure}

After being told that some of the passages they read were co-written with AI, readers expressed significantly more positive perceptions toward the writing compared to the median value ($Z = 3.70$, $p < 0.001$) as well as the writer ($Z = 4.58$, $p < 0.001$) \revision{(See Figure~\ref{fig:part2_post_perception} for the distribution)}. 
This is a surprising result given that previous AIMC work has found readers to hold negative opinions about messages and their writers once they become aware of the message being written with AI assistance~\cite{liu2022will,jakesch2019ai,AIMC-perceptions-of-generated-problematic-replies}. The qualitative responses provided by readers suggest that, unlike interpersonal communication with instrumental or relational purposes, readers in a ``leisurely reading'' setting focus on the writing outcome and their reading experiences, and are positive towards writers experimenting with new technology to improve the outcome. As one reader wrote in their open-text response, ``[using AI writing assistance] could indicate that the human writer is proactive in seeking innovative tools to enhance their creativity and productivity.''

We also observed readers' perceived authorship for work co-written with AI became muddled. In particular, when readers were asked whether they would attribute more credit to the human writers or to AI assistance, there is no statistically significant trend in terms of whether humans or AI should receive more credit for AI-assisted writing ($\beta = 0.18$, $S.E. = 0.19$, $t = 0.97$, $p = 0.339$).
Some readers also acknowledged that using AI tools when writing has become more and more prevalent, and agreed that the general public will eventually be more accepting of this practice.

%\subsection{Future of writing: Writers' desired AI assistance for authentic writing}
\subsection{The Help in Need: Building Feedback Loops for Co-writing with AI}
\label{sec:future_support}

Most writers also believed the public would eventually accept and embrace co-writing with AI.
As such, many of them saw benefits from exploring AI tools and intended to start doing so for writing that ``requires less authenticity'' (P3).
These types of writing were described as grounded more in factual information (e.g., science writing, blog posts for product placement) and less in writers' personal experience and emotional expression.
As a general principle, writers expressed hope that future writing assistance tools could help them preserve their authentic voices in writing, but they preferred support that does \textit{not} come in the form of text production.
Instead, some writers desired more support \textit{before} and \textit{after} the actual writing process, pointing specifically to two areas: (1) constructing fundamental materials for writing (e.g., conducting research, synthesizing relevant information, and assembling related ideas) and (2) actively monitoring, analyzing, and providing feedback on writers' work.

In particular, some writers hoped to build \textbf{``\textit{personalized feedback loops''}}, namely, receiving personalized feedback from AI to improve their work throughout the co-writing processes.
When asked how they would like to select writing samples to personalize their AI writing assistance, nearly all writers preferred using their own work  (only 2 out of 19 writers would include work from others).
Some writers also hoped AI could actively help monitor and evaluate their writing and, most importantly, provide feedback on whether broader audiences would react positively to their work.
Specifically, some writers believed that AI could be most helpful by providing suggestions through an ``algorithmic point of view'' (P13), such as by suggesting words that would be good for SEO optimization. 
But instead of directly receiving and adopting suggestions from the tool, writers expressed preferences for learning from such suggestions and incorporating them into their own writing.
% \placeholder{Mention something about writers' ideal of "good writing"}

\section{Discussion}

\begin{figure}
    \centering
    \includegraphics[width=0.95\linewidth]{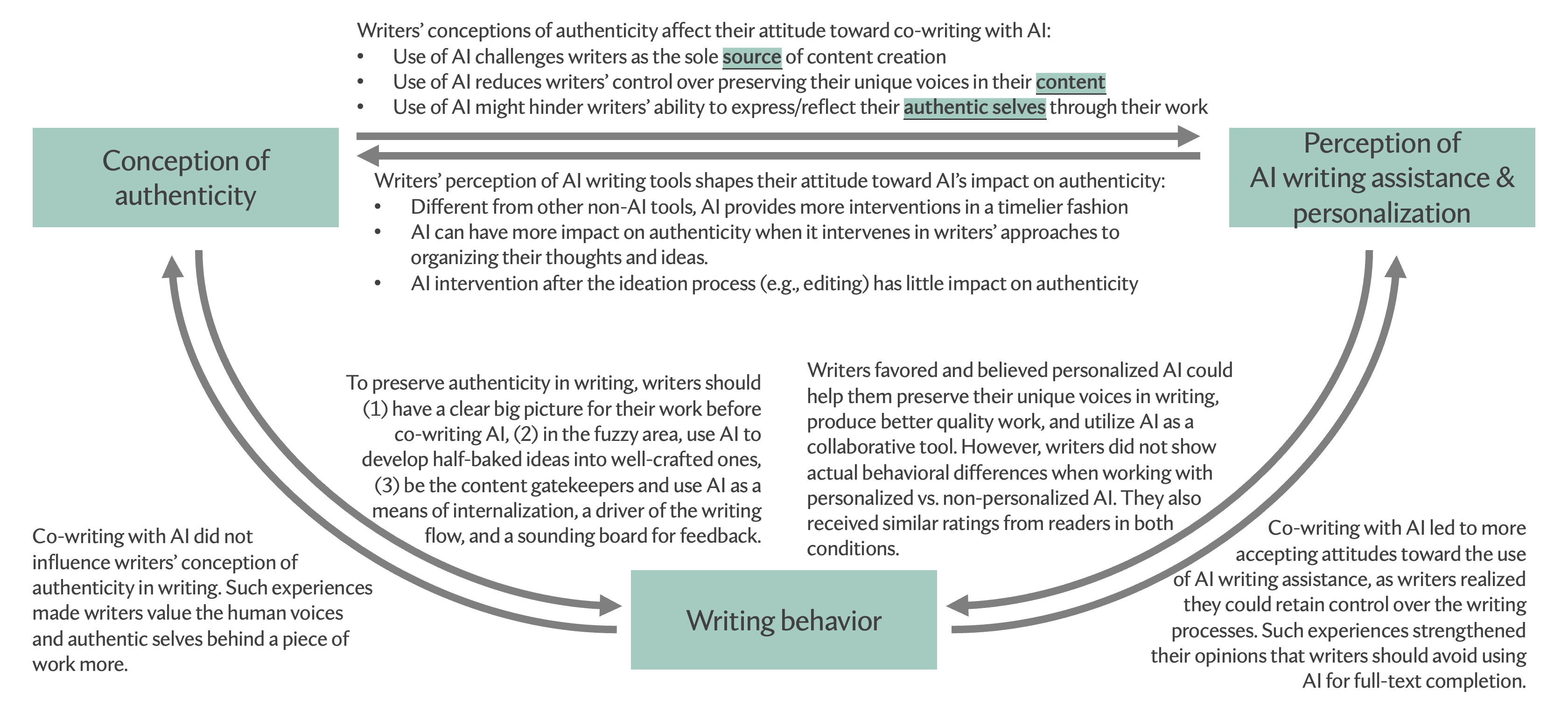}
    \caption{\revision{Summaries of findings from the present research, structured by the relationships between writers' conceptions of authenticity, their perceptions toward AI writing assistance and personalization, and how these attitudes influence their actual behaviors co-writing with AI.}}
    \label{fig:summary_findings}
\end{figure}

% Summary of the present work
In this work, we examined writer-centered conceptions of authenticity in human-AI co-writing.
We explored writers' desired approaches to preserving their authentic voices in writing, and we investigated the effectiveness of personalization in serving this design goal.
Through semi-structured interviews with 19 professional writers, we found writers' conceptions of authenticity tended to focus on their internal experiences---particularly, how they shape and express their authentic selves even before and throughout the writing process.
It is also through their situated practices of co-writing with AI that writers shape and re-shape their perceptions of authenticity --- from persisting writers' producing their work independently as the only way to ensure authenticity to writers' picking up new roles as ``content gatekeepers'' to manage their voices in writing.
\revision{In Figure~\ref{fig:summary_findings}, we summarize findings from the present research and showcase the relationships between writers' conceptions of authenticity, perceptions of AI writing assistance and personalization, and their actual writing behaviors.}
These insights provide not only critical design implications for future AI writing assistance but also practical guides---with an eye toward writers' identities and ethics---for human-AI co-creation in general.
That is, these practical implications shed light on how to create AI tools that help writers complete their jobs without depriving their voices in writing.

\subsection{Revisiting Conceptions of Authenticity in Human-AI Co-Writing}
We begin with consolidating writer-centered definitions of authenticity, which we think is critical to providing an appropriate foundation for designing better AI writing assistance tools.
%Per our writer participants, the two aspects of authenticity does map to existing themes discussed in existing literature. 
Writers' references to the source of work (source authenticity) and writing outcomes (content authenticity) echo the \textit{Source} and \textit{Category} themes of authenticity in existing literature~\cite[e.g.,][]{Auth_Humanity_Guignon_2008, Auth_Humanity_AMA_Lehman_2019, Auth_Humanity_Handler_1986}, although in our study there was less emphasis on content authenticity than in prior work.

Worth noting is that while writer participants' definitions of authenticity shared some similarities with existing conceptions of authenticity used by prior literature, they also highlighted additional nuances that diverge from the existing theoretical frameworks of authenticity.
Considering the idea of \textit{Source} in authenticity: prior literature took a more retrospective view, asking whether one could attribute a piece of work to its creator(s) (e.g.,~\cite{Auth_Humanity_Guignon_2008, Auth_Humanity_AMA_Lehman_2019, Auth_Humanity_Creative_Persona}). Instead, our writer participants focused on various moments during content production: who was taking action and contributing during the actual process of writing? This resembles a more direct way to determine the source of work and its authenticity.

Consider also the concept of \textit{Category} versus \textit{Content authenticity}: while writers in our study also evaluated authenticity in terms of whether the voices and styles of a piece of work represented those of a certain creator(s), they emphasized the professional and economic motivations behind this approach.
Many writers described one's unique writing styles as their ``fingerprints'' and ``brand logo,'' enabling their work to be recognized by broader audiences and their clients.
Conceptually, existing frameworks of authenticity assess work by the degree to which it is representative of a certain literature genre or school of thought (e.g., whether a novel captures the essence of postmodernism)~\cite{Auth_Humanity_Art_Newman_Bloom_2012}.
%For our participants, the unit of categorization went down to the individual level, where authenticity was judged by whether a writer's work echoed their signature pieces.}
However, our participants rarely judged the authenticity of work by genre; they focused solely on whether a piece of text resembles a writer's signature piece. 

Moreover, the writers in our study also centered the concept of authenticity around writers' internal experiences before and throughout the writing process.
Specifically, they highlighted the importance of absorbing, digesting, and internalizing information, knowledge, and lived experiences to nourish their authentic selves and express them in writing.
Ultimately, this becomes their signature, ensuring that a piece of work could not take its eventual shape if done by anyone else.

This last definition of authenticity raised by our writer participants diverges from that of \textit{Value Alignment} in existing theoretical frameworks defining authenticity.
Here, writer participants focused on whether they could genuinely project their internal, living experiences into their writing.
In other words, value alignment implies whether participants could represent their genuine selves and lives in their writing.
This concept of ``value'' is distinct from that in the existing literature, which more often refers to writers' ideology, value systems, and ethics~\cite{Auth_Humanity_Expressive_Authenticity_Lindholm_2013, Auth_Humanity_Art_Newman_Bloom_2012}.
Furthermore, writer participants emphasized that while everyone possesses their own perspectives, emotions, and memories, skills, practices, and iterative processes are required to transform them into writing forms.
%On the other hand, conventional conceptions of value alignment focus on whether writers' views align with what they wrote. Typically, misalignment arises when the writer holds different opinions with other parties, such as their clients.}

Based on these fundamental concepts, we next discuss how the design of human-AI co-writing tools can support writers' authentic selves, source authenticity, and content authenticity.

%\subsection{Designing \textit{for Authenticity} Human-AI Co-Creation Tools through Personalization}
\subsection{Designing AI-writing Tools that Foster \textit{Authenticity} through Personalization}
% The current form of personalization is helpful for creating positive writing experience (which might aid authenticity through their writing processes), but might not directly add benefits to preserving writers' authentic voices
% Instead, AI can provide more support through offering unique "machine point of view" + other forms of personalization
% For writers, authenticity in writing means much more than "sounding like themselves"
% In fact, writers know how to write with and sound like their own authentic writing. Instead, they need other form of support to preserve authenticity

%Our findings highlight the importance of tool design as writers' perception of authenticity as well as whether they can justify authenticity in their writing is rooted in the \textit{practices} of co-writing with AI.
Our findings call attention to how AI writing assistance tools could be designed to support both the different conceptions of authenticity that writers might have and, thus, the \textit{practices} they believe will foster and preserve rather than threaten their sense of authenticity. 
Specifically, our study reveals that authenticity is multifaceted, and the current form of AI assistance (i.e., text suggestion for sentence completion) only contributes to preserving writers' unique styles and voices in writing.
Besides text production, writers seek more diverse support (such as practicing externalizing their internal experiences, receiving feedback, and projecting possible audiences' reactions) to jointly preserve authenticity in their work.

%\subsubsection{\textbf{Absorption comes before production: Supporting \underline{authentic self} through personalizing ``brain food''}} 

%\subsubsection{\textbf{Internalization comes before text production: Supporting \underline{authentic self} through personalizing sources of writing inspiration}} 
\subsubsection{Supporting writers to express their authentic selves through personalized recommendations for sources of writing inspiration} 
% (1) to support the "internal states" aspect of authenticity
%%% (1.a.) AI as internalization tool to help nurish the "real materials" of writing
%%% (1.b.) AI keeping the writing flow
%%% (1.c.) personalization through providing materials for internalization
Our study suggests that authenticity could be supported by enriching the materials that writers learn and draw inspiration from and later on apply to (improve) their writing.
In our interviews, writers highlighted that the most important practice of ``writing'' is, in fact, what they do to nurture themselves with rich information and experiences.
Many writers emphasized how reading plays an important role in improving their writing, and how AI might be able to efficiently support the process of synthesizing and internalizing large amounts of information. 
Thus, providing personalized recommendations for what materials the writers might enjoy or need to engage with can support authenticity by helping writers to enrich their authentic, intellectual selves---which are \textit{the} bases for authentic writing.

%\subsubsection{\textbf{Going beyond the audience's flow: Supporting \underline{source authenticity} through personalizing writers' targets for growth}}
\subsubsection{Providing diverse, early-stage support to help writers establish and accomplish their goals for writing}
% (2) to support the "source" aspect of authenticity
%%% (2.a.) Our study highlights a lot of writers' concerns come from their insecurity toward readers' reaction --> AI to provide more hints of feedback from audiences. this might also support AI to support writers' expressing themselves more freely and genuinely 
%%% --> circle back to AI ghostwriting effect
%%% (2.b.) personalization through personalizing feedback
Across all conversations with writers, we saw how their uncertainty---and oftentimes apprehension---about audiences' possible responses to their work constrained how freely they felt they could express themselves in writing. 
Such concerns also explain some of their hesitations to disclose their use of AI when writing, echoing the AI Ghostwriter Effect~\cite{HAIC_authenticity_AI_Ghostwriter_Effect}.
However, our survey in Part~2 suggests writers' worries about how they and their work would be perceived might not necessarily align with readers' actual attitudes and perceptions of AI-supported writing.
%To more directly tackle the ``source'' aspect of authenticity, we first recognize the need to communicate readers' feedback to writers.
To address writers' concerns about \textit{Source} authenticity (i.e., whether readers would attribute the credit of a piece of work to its author), we first recognize the need to better reflect readers' opinions about credit and authorship to writers.
%Indeed, in our study, writers envisioned AI as a helpful tool to synthesize public opinions and preferences.

%We thus see opportunities for personalization such that writers could get feedback that is tailored to their interests and needs to shape their writing.
Additionally, we see opportunities to build tools that help writers shape their own writing goals and provide feedback accordingly.
%Furthermore, the feedback mechanisms could also place more emphasis on helping writers decide what they want to achieve through their writing. 
For example, personalized AI feedback can help writers decide what types of writer identities to present to their readers and whether they want to highlight their use of new technologies to support their work production processes.  
In particular, offering support during the early stage of idea development (i.e., the ``fuzzy area'' that writer participants referred to) might be more effective in helping writers preserve their authenticity in writing compared to providing support later on during text production. 
It is during this early planning stage that writers often decide what goes into the writing, performing ``content gatekeeping'' that primarily contributes to authenticity, according to many of our writer participants.
%Besides, referencing public interests for goal-setting---instead of following audiences' tastes to craft the actual writing pieces---can also help writers maintain their autonomy during the writing process.

\subsubsection{Supporting content authenticity through personalizing active feedback}
% (3) to support the "writing outcomes" aspect of authenticity
%%% (3.a.) Instead of directly getting content from AI, writers' want to *learn* from AI through feedback loops of good writing
%%% (3.b.) personalization through analysis, metrics, and other machine POV
As described in the idea of building a ``feedback loop of good writing,'' writers in our study were less interested in directly receiving high-quality AI-generated text.
Instead, they were more motivated to learn how to improve and produce good writing through AI tools' feedback.
From our interviews, we also learned that what each writer targets and considers ``good writing'' varies widely from writer to writer.
Therefore, providing customized aids to help writers learn to improve their own writing in the ways they want could be another promising use of personalization.
In our study, writers specifically sought ``machine points of view'' for active monitoring and in-text feedback from the AI tool to improve their writing.
In this regard, enabling writers to select and customize the types of metrics to use for analyzing their writing output might be a possible starting point.

% \placeholder{Mention something about design of for-authenticity tools need to take a more holistic approach; to support the growth of writers from the ground-up, not just the quality of content per se. Because as writers grow, their writing flourishes as well.}
%All in all, the design of for-authenticity AI co-creation tools should take a more holistic approach, 
All in all, we argue that designing AI co-creation tools that preserve authenticity should take a more holistic approach, 
beginning with the perspective that AI assistance need not be narrowly focused on the writing process and output per se. 
Instead, designers and developers of these tools could \textbf{\textit{target the growth of writers as their ultimate design goal}}.
This can be done through AI recommending more diverse reading materials to inspire writers and presenting public feedback for writers to retrospect their own work.
As writers grow, they may seek different topics, content, and styles in their writing, and they may need support not only in improving the quality of their work but also for them to thrive intellectually.
%Therefore, personalizing their ways of writing might offer only temporary aid, and might risk diminishing their sense of ownership and control in writing~\cite{HAIC_authenticity_choice_over_control}.
%Instead, it is critical to diversify sources of inspiration and build more healthy relationships with feedback from broader audiences.
%And most importantly, as writers grow, their writing flourishes as well.
Therefore, instead of personalizing AI text suggestions, providing personalized assistance and feedback that help writers grow can offer more holistic support in the long run.

\subsection{\revision{Open Research Questions for Designing AI Writing Assistance}}

Besides the more concrete design recommendations, as we elaborated in the previous section, writers articulated additional desires and challenges that require future work to explore further the potential of personalized AI tools for addressing these needs. Below, we list four such areas:

\revision{\textbf{\textit{Distribution of work:}} 
Given their concerns about AI's direct threats to source authenticity, writers in our study frequently expressed the perspective that AI writing assistance should not be used to automate text production fully.
In our study, professional writers still desired to conduct most of their writing work. 
Many of them acknowledged performing the writing task themselves helped them improve their work quality and grow as writers.
However, how to effectively distribute the type and load of work between human writers and AI tools---and thus the appropriate level of AI intervention---remains an open question.}

\textbf{\textit{Pacing and temporal adjustment:}}
Throughout our interviews, writer participants described several stages of their writing processes and indicated different types of support needed in each phase.
While writers overall desired control over their writing experiences, we noticed the degree of control needed might also differ from stage to stage.
In this regard, how to adjust the types and frequency of AI support accordingly is a challenging yet important topic for future work.
In particular, personalizing AI assistance through learning individual writers' patterns might be a potential direction for further exploration.

\textbf{\textit{Communication of writers' internal states:}}
Reflecting on writers' living experiences and their internal states remains an under-addressed feature in the design of existing AI writing tools.
Our participants not only suggested these elements as the core of authenticity in human writing, but they could also be used as rich materials for composition.
In this regard, besides designing tools that can support writers during the writing process, additional tools that can help writers to reflect and express their living experiences \textit{before} writing are likewise useful forms of writing assistance.

\textbf{\textit{Writer-reader relationships in the age of human-AI co-creation:}}
% inconsistent view between writers and readers
Taking the findings from Part~1 and Part~2 of our study together, we observe differences between writers' and readers' views.
Despite writers' concerns, readers could not tell when writers adopted AI assistance and reacted rather positively to the idea of writers' experimenting with AI co-writing.
This suggests a possible difference between how writers and readers judge the \textit{Value} of writing output.
For writers, value may be rooted in the hard work that goes into conceiving and producing content; for readers, value may be based on the outcomes of that work, and they
%While writers are concerned about readers' responses to human-AI co-creation, it is important to note that audiences 
do not appear to associate AI-assisted work with a lack of effort.
%In fact, recent work has suggested working with AI assistance could even demand more work from creators~\cite{AIwriting_Tool_integrative_leap_multimodal}.
%The connection between writers and readers remains bonded through the output of writing pieces.
Therefore, devaluation of work may be more likely to arise when writers fail to deliver quality work that meets readers' expectations---which may not require promises about \textit{how} the work is made.

\subsection{What AI Writing Assistance is (Not): Applicability of the Present Insights}
% compare and contrast AIMC vs. AI writing assistance

Providing more clarity to how authenticity is conceptualized in human-AI co-writing also calls us to reconsider %the distinction between AI-assisted writing and other relevant realms, particularly AI-mediated communication (AIMC).
how AI-mediated communication may differ from other AI-assisted writing settings, including the genres we explored in this study.
%By comparing and contrasting the two topics, we further address the generalizability of the present research.
In AIMC, authors adopt AI assistance to produce content for \textit{two-way} communication~\cite{AIMC_Hancock_JCMC_2020}.
This entails delivering messages and meanings to, and expecting reciprocal interaction from, target recipients.
Here, the term ``mediation'' reflects the context where the writing happens, implying AI assistance plays a direct role in altering communication and relationships between writers and readers. %a more participatory role of AI as it possibly alters dialogues between communicators.

On the other hand, while co-writing with AI in more creative settings also produces content, this content is often delivered one way to a larger audience and does not involve a bi-directional message exchange between writers and readers.
Furthermore, our study reveals that the practice of writing extends far beyond crafting the content, and conveying messages to the audience is simply the final step.
Given these distinctions, we expect insights of the present work to apply to comprehending AI's role in facilitating (1) uni-directional presentation of content to a group of individuals and (2) creative practices that nurture the creators and enrich their work, as these practices can extend beyond the content production phase.

\subsection{Limitations and Future Work}
Despite our best efforts and mixed-methods approach to studying this topic, we acknowledge several limitations of the present work as well as the fast-changing landscape and public sentiment in the AI space.
To begin with, data collection for the present work took place from June to October of 2023. 
This is unique timing as the public has been somewhat exposed to the popular trend of generative AI and its use for text production but has not yet moved to complete acceptance and adoption of the new technology.
During this period, new applications of AI writing assistance were also being regularly deployed.
For example, Jasper AI rolled out new features targeting at AI-assisted writing for marketing and branding purposes\footnote{Jasper AI's news press on their launch of new features of AI marketing toolkit: \url{https://www.prnewswire.com/news-releases/jasper-partners-with-google-workspace-webflow-make-and-zapier-to-bring-on-brand-ai-content-across-the-marketing-stack-301911844.html}} soon after we conducted Part~1 of the study. 
Therefore, we also expect writers', readers', and the general public's opinions toward this topic to shift over time.

% Moreover, as abundant media coins the term ``AI hype'' throughout the year of 2023, we suspect responses from both our writers and readers might more or less be affected by popular discussions about generative AI.
Moreover, we expect that our writers' and readers' responses may be affected by the extensive media coverage surrounding generative AI in 2023.
We encourage future work to examine trends across time, capturing and following how writers' and readers' responses change through longitudinal research.
In particular, an interesting topic to explore is whether any of these attitudinal or behavioral changes occur due to advances in new AI applications or due to changes in public regulations, media depictions, or other external factors.

%As both our writer and reader participants carry along unique characteristics, on one hand, these contextual factors add value and richness to the present work.
We acknowledge some findings from the present work that might be related to the unique backgrounds and experiences of our participants. 
As experienced experts, many of our writer participants have developed their own structures and workflows for professional writing.
We noticed their feedback on co-writing with AI might focus more on its potential to facilitate the ideation of content, while they mentioned less about AI's assistance with the structure and composition of writing.
Our reader participants recruited from Reddit might also possess unique characteristics.
Recent surveys showed Reddit users possessed unique social norms, personalities, and demographic breakdowns~\cite{DeCandia_Reddit_2022, PANDORA_Reddit_2021}.
Specifically, Gjurković et al. found Reddit users were rated with particularly high Openness scores in the Big5 personality scales~\cite{PANDORA_Reddit_2021}.
While this can be relevant to our reader participants' accepting attitudes toward writers' use of AI tools for writing, we encourage future work to explore further this relationship and/or conduct studies of relevant topics across multiple platforms.

On the other hand, future work should consider conducting larger-scale studies, surveying both expert and non-expert writers as well as avid and regular audiences. 
This allows researchers to better account for the influences of participants' positionalities on their views toward authenticity. Moreover, our study only examined a limited scope of writer-reader relationships. Specifically, the readers passively and leisurely consumed the writing we provided, with a greater focus on the ``reading experience'' than on seeking out particular writers. Our results may not generalize to readers who actively seek out work from or take a personal interest in a particular writer.

Among the three versions of work produced by our writer participants, the pieces composed without any AI assistance were done without any time constraints.
All writer participants acknowledged that writing under time pressure was a common part of their day-to-day job and that the speed they wrote during the study sessions was at their usual pace. Still, we encourage future work to investigate whether time constraints might influence writers' tendencies to and the ways they engage with AI-writing assistance tools.

Finally, though we supplement writers' perspectives with those of readers, our focus remains on writing as creative work and has not captured views from other stakeholders (e.g., writers' clients) who focus on the potential market value of writing.
We have not yet thoroughly explored how the changing writer economy, to which participants have also hinted as possible influences, might affect writers' decisions and work.
Therefore, we encourage future work to further explore the impact of human-AI co-creation on the monetization of creative work.

\section{Conclusion}
The present work examines writers' and readers' perceptions toward co-writing with AI assistance and its potential impact on the authenticity of writing.
To address these questions, we conducted a two-part research, including a series of in-depth interviews with professional writers and an online study with avid readers.
Despite their hesitation to acknowledge AI co-writing as authentic work at the beginning of the study, nearly all writers recognized authenticity in AI co-writing.
Specifically, they suggested co-writing with personalized AI based on their own writing samples could more effectively support writers in preserving their authentic voices and tones in writing.
From the readers' perspectives, while many of them could tell the difference between a writer's independent work and AI-assisted work, they appreciated writers experimenting with new technologies to perform their writing jobs and expressed interest in reading content co-written with AI.
Together, the present work makes three contributions: 
(1) We synthesize some writer-centered conceptions of authenticity in the age of human-AI collaboration. 
(2) We provide insights into how to support writers in preserving their authenticity in writing. 
(3) We reflect on readers' attitudes toward reading AI-co-writing work.
Together, these insights offer practical implications for designing future AI co-writing tools.

\begin{acks}
We thank colleagues from the Microsoft FATE Team for their continuous feedback as we developed and conducted the study. Among them, we want to give a special shout-out to the 2023 Summer Research Intern cohort who participated in our pilot studies to help us refine our study protocol.
\end{acks}


%%
%% The next two lines define the bibliography style to be used, and the bibliography file.
\bibliographystyle{ACM-Reference-Format}
\bibliography{paper_ref, ref/AI_writing_tool, ref/AIMC, ref/authenticity_HAIC, ref/authenticity_humanity, ref/method, ref/personalization}

\appendix

%%%%%%%%%%%%%%%% Part 1 Pre-Survey %%%%%%%%%%%%%%%%
% \newpage
\section*{Appendix}
\section{Pre-study Survey from Part~1 of the Study}
\label{appendix:pre-survey}

\begin{enumerate}
    \item How would you describe your own writing? What makes your writing unique? Can you point out some unique characteristics (e.g., tones, voices, styles, features, etc.) in your writing? [Open text response]
    \item Do you have any prior experience co-writing with any AI tool (ChatGPT, Grammerly, Jasper AI, Peppertype, Anyword, etc.)? If yes, please elaborate. [Open text response]
    \item Do you use other tools, resources, materials, etc. to support your writing process? If yes, please describe them. [Open text response]
    \item Is using these other writing support tools/resources any different from using an AI writing assistance tool? If yes, how so? (Please put n/a in the field if you do not use other forms of writing support.) [Open text response]
    \item How would you define authenticity in writing? What does it mean to you as a writer?
    \item Please submit a short piece of text (~ 200 words) that you have written in the past. The text should represent your unique, authentic "voice" in writing.
\end{enumerate}

%%%%%%%%%%%%%%%% Part 1 Interview %%%%%%%%%%%%%%%%
%\newpage
\section{Interview Protocol for Part~1 of the Study}
\label{appendix:interview_protocol}

\subsection{Intro/warm-up questions}
During the intro session, we asked participants to elaborate further on their responses in the pre-study questionnaires, particularly their perspectives on authenticity in writing.

\begin{itemize}
    \item What are the unique characteristics (tones, phrases, styles, voices, etc.) that make your writing unique?
    \item What is your experiences co-writing with AI tools?
    \item What is your definition of authenticity in writing?
    \item Based on your own definition of authenticity, would you consider co-writing with AI as “authentic”?
    \item Would you want to preserve your unique characteristics in writing when you co-write with AI?
    \item How is using other non-AI tools to support writing similar to / different from using AI tools to do so?
    \item What does "good writing" mean for you?
\end{itemize}

\subsection{Questions after each AI co-writing session}
After each AI co-writing session, we asked participants the following questions. We asked the same set of questions for the personalized and non-personalized session.

\begin{itemize}
    \item What was your overall experience co-writing with the AI tool?
    \item Does co-writing with AI have an effect on your writing \textit{outcome}?
    \item Does co-writing with AI have an effect on your writing \textit{process}?
    \item How did you feel about being affected by the AI writing tool? 
    \item After co-writing with AI, does that change how you think about authenticity in writing?
    \item After co-writing with AI, do you consider co-writing with AI as authentic writing?
    \item Does co-writing with AI affect your authenticity in writing? 
    \item Did you do anything to prevent AI from affecting your authenticity in writing?
\end{itemize}

\subsection{Final Thoughts}
After participants completed both AI co-writing sessions and responded to those above-mentioned questions, we then revealed to them the difference between the two sessions (i.e., personalized or not) and asked the following questions:

\begin{itemize}
    \item Does the personalized AI tool pick up your unique voice, tone, style, etc. in writing?
    \item Do you prefer co-writing with a personalized or a generic AI tool?
    \item Is this the right amount of “personalization”?
    \item How will you select writing samples to personalize your AI writing tool?
    \item Does co-writing with a personalized AI produce more authentic work?
    \item Does co-writing with AI change how you think about authenticity in writing?
    \item How might your readers react to the idea of you co-writing with an AI?
    \item Do you have any concerns about writing with a personalized AI tool?
    \item Are there other ways that AI can support authenticity in writing?
    \item Are there other ways that AI can support creativity and good writing?
\end{itemize}

%%%%%%%%%%%%%%%% Part 1 Codebook %%%%%%%%%%%%%%%%
\newpage
\section{Codebook of Data Analysis from Part~1 the Study}
\label{appendix:codebook}

% \newpage
\begin{table}[h!]
\begin{minipage}{0.48\textwidth}
    %\centering
    \resizebox{0.99\textwidth}{!}{
    \begin{tabular}{p{0.9\textwidth}|p{0.1\textwidth}}
    \bottomrule
    \rowcolor{lightgray!20!}
    \multicolumn{2}{c}{Writer-centered definition of authenticity} \\
    \hline
        Theme/Code & Count \\
        \hline
        \textbf{Content authenticity} &  \\
        $\bullet$ Consistency & 4 \\
        $\bullet$ Word choice & 8 \\
        \hline
        \textbf{Source authenticity} & \\
        $\bullet$ Source of content & 9 \\
        $\bullet$ Original source & 9 \\
        \hline
        \textbf{Authentic self} & \\
        \hspace{3mm} $\bullet$ Writer's identity &  \\
        \hspace{7mm} $-$ Living experience & 16 \\
        \hspace{7mm} $-$ Writers' background and names & 12 \\
        \hspace{7mm} $-$ Personal story & 9 \\
        \hspace{7mm} $-$ Emotional value and expression & 8 \\
        \hspace{7mm} $-$ Personal value and importance & 4 \\
        \hspace{3mm} $\bullet$ Justification for one's work & 7 \\
        \hspace{3mm} $\bullet$ Expression of self & 5 \\
        \hspace{3mm} $\bullet$ Context of writing & 3 \\
        \hline
        \rowcolor{lightgray!20!}
        \multicolumn{2}{c}{Writing practices to preserve authentic voices} \\
        \hline
        \textbf{Starting point of writing} & \\
        \hspace{3mm} $\bullet$ Direction of writing & 10 \\
        \hspace{3mm} $\bullet$ Clear vision & 5 \\
        \hline
        \textbf{Stages and timing for AI assistance} & \\
        \hspace{3mm} $\bullet$ Early ideation & 15 \\
        \hspace{3mm} $\bullet$ Vague idea and fuzzy area & 8 \\
        \hspace{3mm} $\bullet$ Planning for writing & 2 \\
        \hline
        \textbf{Content gate keeping} & \\
        \hspace{3mm} $\bullet$ Portion of contribution & 17 \\
        \hspace{3mm} $\bullet$ Selection of AI suggestions & 16 \\
        \hspace{3mm} $\bullet$ Revision of AI suggestions & 13 \\
        \hspace{3mm} $\bullet$ Control for content & 11 \\
        \hspace{3mm} $\bullet$ Cut content & 6 \\
        \hspace{3mm} $\bullet$ Balance human and AI input & 5 \\
        \hspace{3mm} $\bullet$ Blending and combining AI suggestions & 5 \\
        \hline
        \textbf{Opportunities for AI to support authenticity} & \\
        \hspace{3mm} $\bullet$ AI as internalization tool & \\
        \hspace{7mm} $-$ Internalization & 14 \\
        \hspace{10mm} $\cdot$ Learn from other writers & 2 \\
        \hspace{10mm} $\cdot$ Reading for writing & 2 \\
        \hspace{10mm} $\cdot$ Gathering information and references & 8 \\
        \hspace{10mm} $\cdot$ Expedite internationalization & 2 \\
        \hspace{7mm} $-$ Finding inspiration from the real world & 10 \\
        \hspace{7mm} $-$ Capture internal states & 3 \\
        \hspace{3mm} $\bullet$ AI as sounding board & \\
        \hspace{7mm} $-$ Communication & 4 \\
        \hspace{7mm} $-$ Gauge audience's responses & 14 \\
        \hspace{7mm} $-$ Understanding public expectations & 6 \\
        \bottomrule
    \end{tabular}}
\end{minipage} \hfill
\begin{minipage}{0.48\textwidth}
    %\centering
    \resizebox{0.99\textwidth}{!}{
    \begin{tabular}{p{0.9\textwidth}|p{0.1\textwidth}}
    \bottomrule
        Theme/Code & Count \\
        \hline
        \textbf{Opportunities for AI to support authenticity (cont'd)} & \\
        \hspace{3mm} $\bullet$ AI as driver of flow & \\
        \hspace{7mm} $-$ Content continuation & 13 \\
        \hspace{7mm} $-$ Remove writing blocks & 9 \\
        \hspace{7mm} $-$ Flow experience & 8 \\
        \hspace{7mm} $-$ Keep up with productivity & 8 \\
        \hspace{7mm} $-$ Continue with the momentum & 8 \\
        \hspace{7mm} $-$ Jumping points & 6 \\
        \hspace{7mm} $-$ Bridge ideas & 4 \\
        \hline
        \rowcolor{lightgray!20!}
        \multicolumn{2}{c}{Personalization as double-edged sword} \\
        \hline
        \textbf{Preferences for personalization} & \\
        \hspace{3mm} $\bullet$ Fit with one's tone & 8 \\
        \hspace{3mm} $\bullet$ Collaboration & 5 \\
        \hspace{3mm} $\bullet$ Simulate one's voice and style & 5 \\
        \hline
        \textbf{Concerns for personalization} & \\
        \hspace{3mm} $\bullet$ Imitation of styles & 10 \\
        \hspace{3mm} $\bullet$ Reliance on AI & 8 \\
        \hspace{3mm} $\bullet$ Diversity in language & 5 \\
        \hline 
        \rowcolor{lightgray!20!}
        \multicolumn{2}{c}{Writer-reader relationship} \\
        \hline
        \textbf{Concerns for work devaluation} & \\
        \hspace{3mm} $\bullet$ Work devaluation & 13 \\
        \hspace{7mm} $-$ Opinions from clients & 3 \\
        \hspace{7mm} $-$ Opinions from non-experts & 3\\
        \hspace{7mm} $-$ Writer economy & 5 \\
        \hspace{7mm} $-$ Work ethics & 2 \\
        \hspace{3mm} $\bullet$ Copyright and plagiarism & 4 \\
        \textbf{Connection with readers} & 11 \\
        \textbf{Acceptance of AI} & 10 \\
        \hline 
        \rowcolor{lightgray!20!}
        \multicolumn{2}{c}{Desired writing assistance} \\
        \hline
        \textbf{Assistance beyond writing} & \\
        \hspace{3mm} $\bullet$ Improvement & 4 \\
        \hspace{3mm} $\bullet$ Presentation & 4 \\
        \hspace{3mm} $\bullet$ Organization and structure & 3 \\
        \hline
        \textbf{Personalzied feedback loop} & \\
        \hspace{3mm} $\bullet$ Selection of writing samples & 19 \\
        \hspace{7mm} $-$ Learning/training materials & 3 \\
        \hspace{3mm} $\bullet$ Good writing & 17 \\
        \hspace{3mm} $\bullet$ Feedback loop & 10 \\
        \hspace{3mm} $\bullet$ Machine point-of-view & 8 \\
        \hspace{3mm} $\bullet$ Active analyzing and monitoring & 6 \\
        \hspace{3mm} $\bullet$ Evaluation and comparison & 6 \\
        \hspace{3mm} $\bullet$ Provide options and alternatives & 5 \\
        \hspace{3mm} $\bullet$ Prioritize audience's perspectives & 5\\
        \bottomrule
    \end{tabular}}
\end{minipage}
\end{table}

%%%%%%%%%%%%%%%% Part 1 Writing log %%%%%%%%%%%%%%%%
\newpage
\section{Writers' Behavioral Data Recorded from Writing Logs in Part~1}
\label{appendix:writing_log}

We used the CoAuthor interface to record their writing logs and their final writing output.
Writers' behavioral data recorded through the writing logs include (1) the frequency and timing of their requesting AI suggestions, (2) the frequency and timing of them accepting and/or rejecting AI suggestions, (3) the AI suggestions (in text) provided at each of their requests, (4) the AI suggestions (in text) accepted, if any, and (5) the text inserted by writers. 
Below, we reported the frequency of writers' requesting AI suggestions and the portion (by \%) of suggestions accepted.

\begin{figure}[h!]
    \centering
    \includegraphics[width=0.9\textwidth]{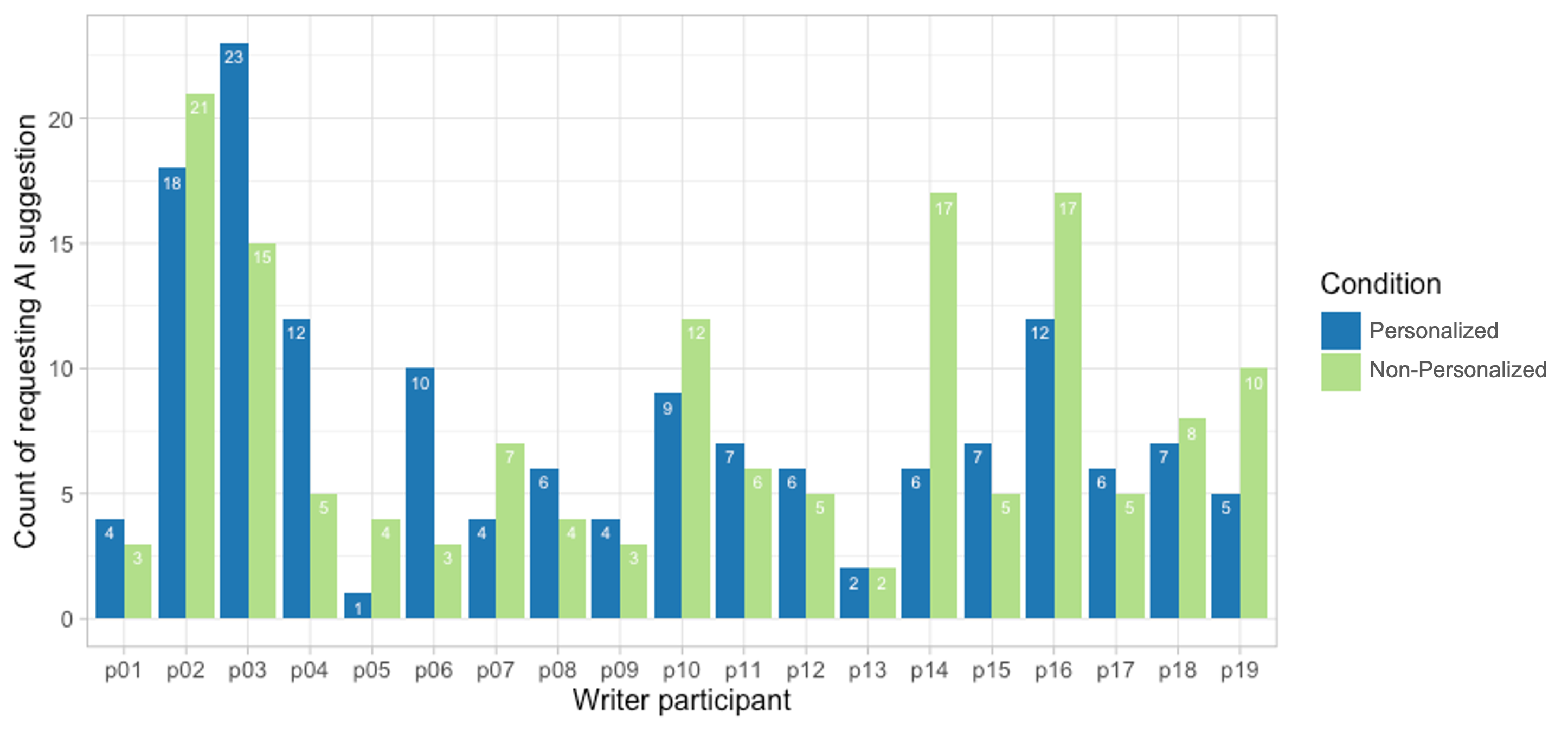}
    \caption{Number of times each writer participant pulled up AI assistance when writing with the personalized vs. non-personalized tools.}
    \label{fig:part1_get}
\end{figure}

\begin{figure}[h!]
    \centering
    \includegraphics[width=0.9\textwidth]{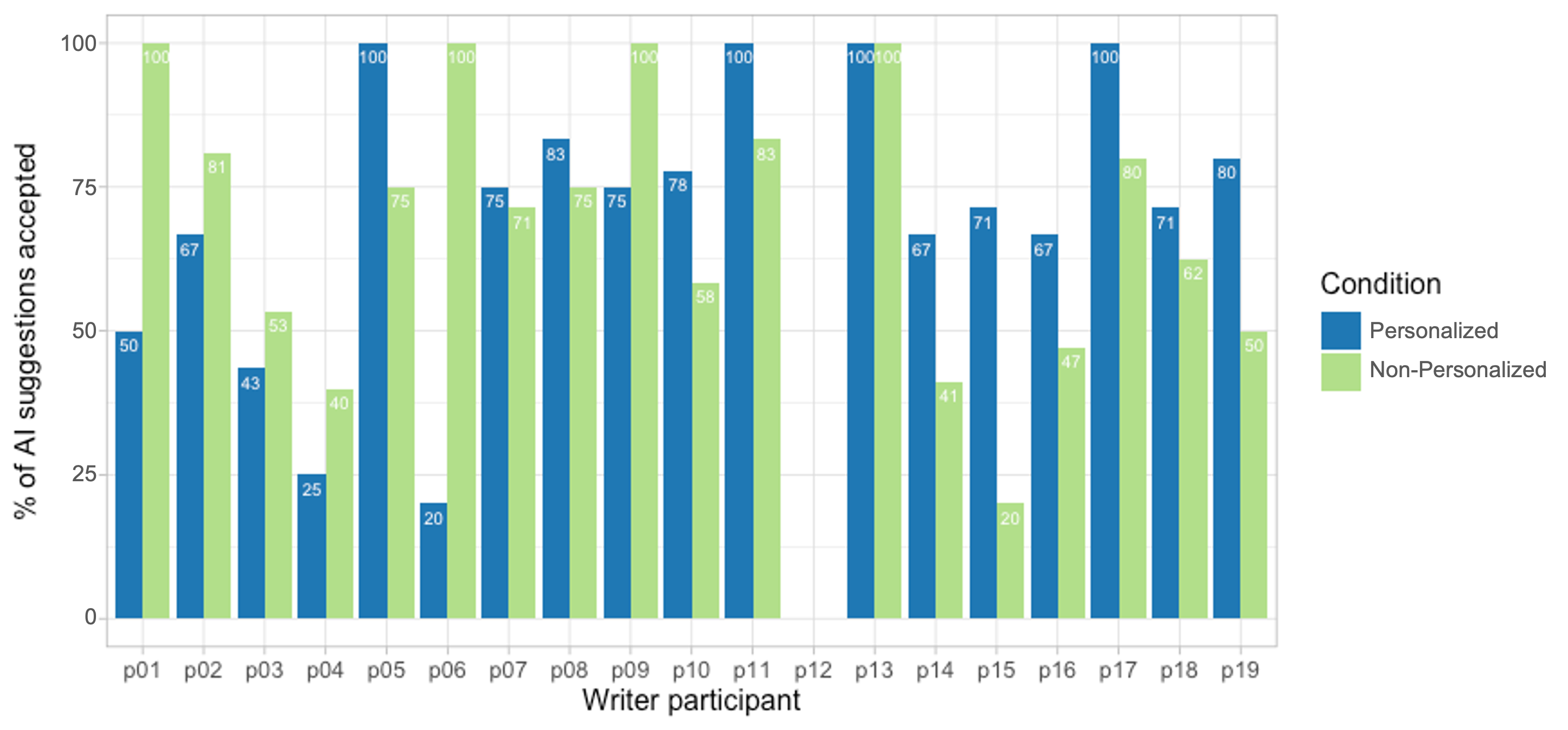}
    \caption{Rate (\%) of AI suggestions accepted by each writer participant when writing with the personalized vs. non-personalized tools.}
    \label{fig:enter-label}
\end{figure}

%%%%%%%%%%%%%%%% Part 2 Subreddit %%%%%%%%%%%%%%%%
\newpage
\section{Subreddits for Participant Recruitment in Part~2}
\label{appendix:subreddit}

General groups that discuss reading and books:
\begin{itemize}
    \item r/suggestmeabook
    \item r/literature
    \item r/52book
    \item r/books
    \item r/bookdiscussion
    \item r/currentlyreading
    \item r/goodreads
    \item r/whattoreadwhen
\end{itemize}

Groups that discuss specific literature genres:
\begin{itemize}
    \item r/Poetry
    \item r/Fantasy
    \item r/RomanceBooks
    \item r/scifi
\end{itemize}

%%%%%%%%%%%%%%%% Part 2 Participant Reading Genre %%%%%%%%%%%%%%%%
\section{Frequently read literature genres by reader participants in Part 2}
\label{appendix:reader_genre}

\revision{Below enumerates a list of literature genres that our reader participants at least once per week. Note that each participant might read more than one genre on a regular basis.
\begin{itemize}
    \item Fantasy: 67.03\%
    \item Science fiction: 66.49\%
    \item Romance: 64.32\%
    \item Contemporary fiction: 63.24\%
    \item Thrillers and horror: 62.70\%
    \item Mystery: 22.16\%
    \item Historical fiction: 20.00\%
    \item Inspiration and self-help: 17.84\%
    \item Biography, autobiography, and memoir: 6.49\%
    \item Poetry: 2.16\%
    \item Drama and screenplay: 1.62\%
\end{itemize}
}

%\input{table/table_reader_genre}

%%%%%%%%%%%%%%%% Part 2 Survey %%%%%%%%%%%%%%%%
%\newpage
\section{Full Survey from Part~2 of the Study}
\label{appendix:part2_survey}

The beginning of the questionnaire includes an informed consent form.
Participants needed to read and consent to participate before they could proceed to the survey questions.

\subsection{Opinions about AI writing assistance}

\begin{itemize}
    \item Are you familiar with the recent trend in generative AI, such as people can use tools, like ChatGPT, to generate text for writing? [5-point Likert scales; Not familiar at all = 1 $\leftrightarrow$ Extremely familiar = 5]
    \item Please describe what you know about the recent trend in generative AI. [Open text response]
    \item Are you interested in reading work co-written by human writers and AI? [5-point Likert scales; Not interested at all = 1 $\leftrightarrow$ Extremely interested = 5]
    \item Do you have any concern about writers who might use AI to help them write? [Open text response]
    \item As a reader, what are the qualities of "good writing" that you most appreciate? [Open text response]
    \item As a reader, what does authenticity in writing mean to you? [Open text response]
    \item Based on your definition of authenticity, do you consider a piece of work co-written by a human and an AI writing assistance tool as authentic writing? [5-point Likert scales; Definitely not authentic writing = 1 $\leftrightarrow$ Definitely as authentic writing = 5]
    \item Please describe your reason(s) for the question above. [Open text response]
\end{itemize}

\subsection{Reading writing samples}
Each participant read three writing samples from one writer, including the writer's solo work, the passage they co-wrote with \textit{personalized} AI assistance, and the passage they co-wrote with \textit{non-personalized} AI assistance. The order of these three passages are presented in a randomized order. Immediately after reaching each passage, each participant uses the following scales to rate the writing:

\begin{itemize}
    \item How much do you \underline{like} the writing? [5-point Likert scales; Not at all = 1 $\leftrightarrow$ A great deal = 5]
    \item How much do you \underline{enjoy} the writing? [5-point Likert scales; Not at all = 1 $\leftrightarrow$ A great deal = 5]
    \item How much do you \underline{creative} the writing? [5-point Likert scales; Not at all = 1 $\leftrightarrow$ A great deal = 5]
\end{itemize}

\subsection{Comparing all writing samples}
We presented the three pieces of writing side by side on the same page. We randomly assigned labels (Writing A, Writing B, and Writing C) to the passages. We gave participants the following instructions and asked them to compare the three using 5-point Likert scales in a matrix layout:

Instruction for evaluation: \textit{``Below are the three pieces of writing you read in the previous pages. These pieces were either written independently by a human author or co-written by the same human writer and an AI writing assistance tool. Which piece(s) of writing was written by a human writer independently and which was co-written with an AI writing assistance tool?''}

\begin{table}[h!]
    \centering
    \resizebox{.99\textwidth}{!}{
    \begin{tabular}{p{0.1\textwidth} | p{0.2\textwidth} | p{0.2\textwidth} | p{0.15\textwidth} | p{0.22\textwidth} | p{0.23\textwidth}}
    \toprule
         & Definitely co-written with AI & Probably co-written with AI & No idea & Probably written independently by a human writer & Definitely written independently by a human writer \\
         \hline
         Writing A & & & & & \\
         \hline
         Writing B & & & & & \\
         \hline
         Writing C & & & & & \\
    \bottomrule
    \end{tabular}}
    \label{tab:my_label}
\end{table}

\subsection{Comparing personalized vs. non-personalized writing samples}
We again presented the three pieces of writing side by side on the same page. We gave participants the following instructions and asked them to compare the personalized and non-personalized writing pieces to the writers' solo work using 5-point Likert scales in a matrix layout and open-text:

Instruction for evaluation: \textit{``Writing [\# of writers' solo work] was written independently by a human writer. Writing [\# of work co-written with personalized AI] and Writing [\# of work co-written with non-personalized AI] were co-written by the human writer and an AI writing assistance tool. Compared these two pieces to the text written independently by the author (Writing [\# writer's solo work]), and answer the following questions.''}

\begin{itemize}
    \item Compared to the text written independently by the author, to what extent do you think the co-written text preserves the authentic voice of the author?
\end{itemize}

\begin{table}[h!]
    \centering
    \resizebox{.99\textwidth}{!}{
    \begin{tabular}{p{0.5\textwidth} | p{0.1\textwidth} | p{0.1\textwidth} | p{0.2\textwidth} | p{0.1\textwidth} | p{0.1\textwidth}}
    \toprule
         & None at all & A little & A moderate amount & A lot & A great deal \\
         \hline
         Writing [\# of work co-written with \textbf{personalized} AI] & & & & & \\
         \hline
         Writing [\# of work co-written with \textbf{non-personalized} AI] & & & & & \\
    \bottomrule
    \end{tabular}}
    \label{tab:my_label}
\end{table}

\begin{itemize}
    \item From Writing [\# personalized], please copy the part(s) of writing that preserves the authentic tone and voice of the writer, if at all. [Open text response]
    \item From Writing [\# non-personalized], please copy the part(s) of writing that preserves the authentic tone and voice of the writer, if at all. [Open text response]
    \item Comparing text co-written with AI (Writing [\# personalized] and Writing [\# non-personalized]) to the author's independent writing (Writing [\# solo]), who do you think should own credits for the work?
\end{itemize}

\begin{table}[h!]
    \centering
    \resizebox{.99\textwidth}{!}{
    \begin{tabular}{p{0.5\textwidth} | p{0.12\textwidth} | p{0.12\textwidth} | p{0.12\textwidth} | p{0.12\textwidth} | p{0.12\textwidth}}
    \toprule
         & The \textbf{AI} definitely owns more credits & The \textbf{AI} probably owns more credits & Both own equal credits & The \textbf{human writer} probably owns more credits & The \textbf{human writer} definitely owns more credits \\
         \hline
         Writing [\# of work co-written with \textbf{personalized} AI] & & & & & \\
         \hline
         Writing [\# of work co-written with \textbf{non-personalized} AI] & & & & & \\
    \bottomrule
    \end{tabular}}
    \label{tab:my_label}
\end{table}

\begin{itemize}
    \item Comparing text co-written with AI (Writing [\# personalized] and Writing [\# non-personalized]) to the author's independent writing (Writing [\# solo]), who do you think should claim authorship for the work?
\end{itemize}

\begin{table}[h!]
    \centering
    \resizebox{.99\textwidth}{!}{
    \begin{tabular}{p{0.5\textwidth} | p{0.12\textwidth} | p{0.12\textwidth} | p{0.12\textwidth} | p{0.12\textwidth} | p{0.12\textwidth}}
    \toprule
         & The \textbf{AI} should definitely claim primary authorship & The \textbf{AI} should probably claim primary authorship & Both should claim equal authorship & The \textbf{human writer} should probably claim primary authorship & The \textbf{human writer} should definitely claim primary authorship \\
         \hline
         Writing [\# of work co-written with \textbf{personalized} AI] & & & & & \\
         \hline
         Writing [\# of work co-written with \textbf{non-personalized} AI] & & & & & \\
    \bottomrule
    \end{tabular}}
    \label{tab:my_label}
\end{table}

\subsection{Final thoughts after reading AI-assisted writing}

\begin{itemize}
    \item Overall, how do you feel about an author using an AI writing tool to facilitate their writing processes? [Open text response]
    \item Does knowing that a piece of text is co-written with an AI tool affect your perception of the writing? [5-point Likert scales; Much more negative = 1 $\leftrightarrow$ Much more positive = 5]
    \item Please elaborate on your reason(s) for the question above. [Open text response]
    \item Does knowing that a piece of text is co-written with an AI tool affect how you would appreciate and evaluate the writing? [5-point Likert scales; Not at all = 1 $\leftrightarrow$ A great deal = 5]
    \item Please elaborate on your reason(s) for the question above. [Open text response]
    \item Does knowing that a piece of text is co-written with an AI tool affect your perception of the human writer? [5-point Likert scales; Much more negative = 1 $\leftrightarrow$ Much more positive = 5]
    \item Please elaborate on your reason(s) for the question above. [Open text response]
\end{itemize}

\end{document}